\documentclass[graybox,natbib,nosecnum]{svmult}
\bibpunct{(}{)}{;}{a}{}{,} 

\pdfoutput=1   

\usepackage{mathptmx}       
\usepackage{helvet}         
\usepackage{courier}        
\usepackage{type1cm}        

\usepackage{makeidx}         
\usepackage{graphicx}        
\usepackage{multicol}        
\usepackage[bottom]{footmisc}
\usepackage[normalem]{ulem}	
\usepackage{hyperref}  

\usepackage{soul}   

\def\ms{\hbox{m\,s$^{-1}$}}
\def\kms{\hbox{km\,s$^{-1}$}}
\def\msun{\hbox{${\rm M}_{\odot}$}}
\def\mearth{\hbox{${\rm M}_{\oplus}$}}
\def\rearth{\hbox{${\rm R}_{\oplus}$}}
\def\arcsec{\hbox{$^{\prime\prime}$}}
\def\degr{\hbox{$^\circ$}}

\makeindex             

\begin{document}

\title*{SPIRou: a nIR spectropolarimeter / high-precision velocimeter for the CFHT} 
\titlerunning{SPIRou: a nIR spectropolarimeter / velocimeter for the CFHT} 
\author{JF~Donati, D~Kouach, M~Lacombe, S~Baratchart, R~Doyon, X~Delfosse, E~Artigau, C~Moutou, G~H\'ebrard, F~Bouchy, J~Bouvier, S~Alencar, L~Saddlemyer, 
L~Par\`es, P~Rabou, Y~Micheau, F~Dolon, G~Barrick, O~Hernandez, SY~Wang, V~Reshetov, N~Striebig, Z~Challita, A~Carmona, S~Tibault, E~Martioli, 
P~Figueira, I~Boisse, F~Pepe \& the SPIRou team} 
\authorrunning{JF~Donati \& the SPIRou team} 
\institute{JF Donati \at IRAP/OMP, Toulouse, France.  \email{jean-francois.donati@irap.omp.eu} 
\and D~Kouach, M~Lacombe, S~Baratchart, L~Par\`es, Y~Micheau, N~Striebig, Z~Challita, A~Carmona, IRAP/OMP, France  
\and R~Doyon, E~Artigau, O~Hernandez, S~Tibault, UdeM/UL, Canada 
\and X~Delfosse, J~Bouvier, P~Rabou, IPAG, France
\and C~Moutou, G~Barrick, CFHT, Hawaii 
\and G~H\'ebrard, IAP/IdF, France 
\and F~Bouchy, I~Boisse, F~Dolon, LAM/OHP, France
\and S~Alencar, UFMG, Brazil 
\and L~Saddlemyer, V~Reshetov, NRC-H, Canada 
\and SY~Wang, ASIAA, Taiwan 
\and E~Martioli, LNA, Brazil
\and P~Figueira, CAUP, Portugal
\and F~Pepe, OG, Switzerland} 
%
%
\maketitle

\abstract{SPIRou is a near-infrared (nIR) spectropolarimeter / velocimeter for the Canada-France-Hawaii Telescope (CFHT), 
that will focus on two forefront science topics, (i) the quest for habitable Earth-like planets around nearby M stars, 
and (ii) the study of low-mass star/planet formation in the presence of magnetic fields.  
SPIRou will also efficiently tackle many key programmes beyond these two main goals, from weather patterns on brown dwarfs 
to Solar-System planet and exoplanet atmospheres. 
SPIRou will cover a wide spectral domain in a single exposure (0.98-2.44~$\mu$m) at a resolving power of 70~K, yielding 
unpolarized and polarized spectra of low-mass stars with a 15\% average throughput at a radial velocity (RV) precision of 1~\ms.  
It consists of a Cassegrain unit mounted at the Cassegrain focus of CFHT and featuring an achromatic polarimeter, coupled to a 
cryogenic spectrograph cooled down at 80~K through a fluoride fiber link.  
SPIRou is currently integrated at IRAP/OMP and will be mounted at CFHT in 2017 Q4 for a first light scheduled in late 2017.  
Science operation is predicted to begin in 2018 S2, allowing many fruitful synergies with major ground and space instruments such 
as the JWST, TESS, ALMA and later-on PLATO and the ELT.}

\section{Introduction}

Detecting and characterizing exoplanets, especially Earth-like ones located at the right distance from their host stars 
to lie in the habitable zone (HZ, where liquid water can pool at the planet surface), stands as one of the most exciting areas 
of modern astronomy and comes as an obvious milestone in our quest to understand the emergence of life \citep{Gaidos07}.  
High-precision velocimetry, measuring RVs of stars and the periodic fluctuations that probe the presence of orbiting bodies, 
is currently the most reliable way to achieve this goal;  in particular, velocimetry allows one to validate candidate planets 
detected with transit surveys (with, e.g., CoRoT, Kepler, TESS, and later-on PLATO), and to estimate the densities and study 
the bulk composition of the detected planets from their masses and radii \citep{Lissauer14}.  

M dwarfs are key targets for this quest;  beyond largely dominating the population of the solar neighborhood, they feature many 
low-mass planets \citep{Dressing15, Gaidos16} and render HZ planets far easier to detect by shrinking the size of their HZs 
(thereby boosting RV wobbles and reducing orbital periods).  Their monitoring with existing velocimeters like HARPS on the 3.6m 
ESO telescope \citep{Rupprecht04} is however tricky, especially for the coolest ones, given their intrinsic faintness 
at visible wavelengths, preventing a deep-enough 
exploration to  detect significant samples of HZ Earth-like planets \citep{Bonfils13}.  Moreover, M dwarfs are notorious for their 
magnetic activity, generating spurious RV signals (activity jitter) that can hamper planet detectability \citep{Newton16, Hebrard16}.  

Modeling the activity of M dwarfs and the underlying magnetic fields is thus crucial for filtering out the RV jitter and for 
maximizing the efficiency at detecting low-mass planets \citep{Hebrard16}.  Magnetic fields of low-mass stars are also expected 
to have a major impact on the evolution of close-in planets \citep{Strugarek15} as well as on their habitability \citep{Gudel14, 
Vidotto13}.  Allowing one to detect and model large-scale fields of active stars, spectropolarimetry comes as the 
ideal complement to precision velocimetry, making it possible not only to maximise the efficiency of planet detection, but also 
to characterise the impact of magnetic activity on the habitability of the detected close-in planets.  

Investigating star/planet formation comes as the logical complement to studying exoplanetary systems of M dwarfs.  Magnetic fields 
are known to have a major impact at the early stages of the life of low-mass stars and their planets, as they form from collapsing 
dense pre-stellar cores that progressively flatten into large-scale magnetized accretion discs and eventually settle as young suns 
orbited by planetary systems \citep{Andre09}.  In this overall picture, the pre-main-sequence (PMS) phases, in which central protostars 
feed from surrounding planet-forming accretion discs, are crucial for our understanding of how worlds like our Solar System form.  
Following a phase where they massively accrete from their discs (as class-I protostars, aged 0.1-0.5 Myr) while still embedded in 
dust cocoons, newly formed protostars progressively grow bright enough to clear out their dust envelopes (at ages 0.5-10 Myr), 
becoming classical T-Tauri stars (cTTSs) when still accreting from their planet-forming discs, then weak-line T-Tauri stars (wTTSs) 
once they have mostly exhausted their discs.  These steps are key for benchmarking star/planet formation.  

\begin{figure}[t]
\includegraphics[scale=0.68]{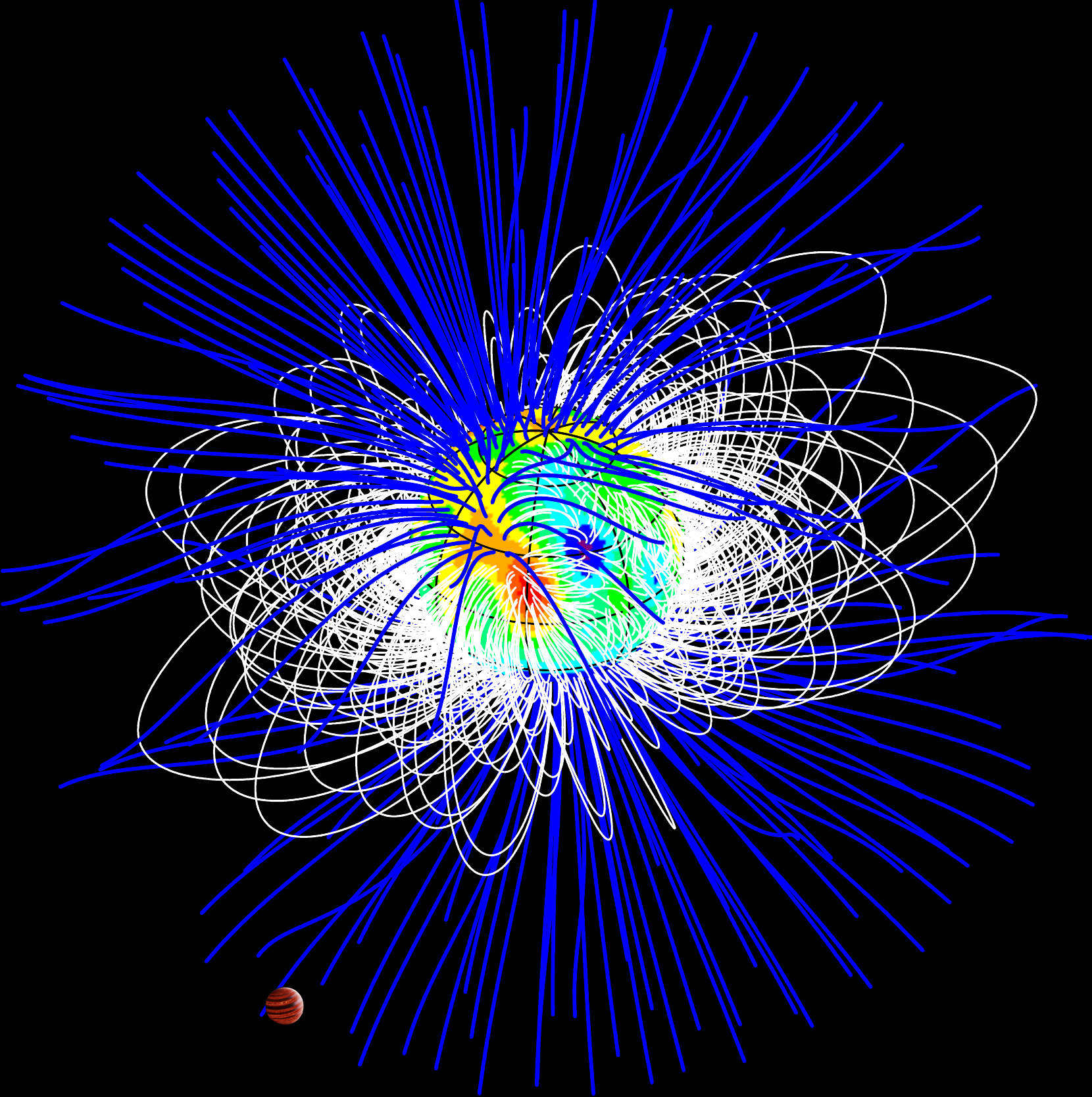}
\caption{Artist view of the hot Jupiter recently detected around the wTTS V830~Tau, orbiting in the large-scale magnetic field of its 
host star.  The field topology is reconstructed using tomographic imaging on a phase-resolved spectropolarimetric data set of V830~Tau
\citep{Donati17}.} 
\label{fig:lsfp}
\end{figure}

Spectropolarimetry is the ideal tool for constraining the large-scale field topologies of PMS stars and their accretion discs, and 
thereby quantitatively assess the impact of magnetic fields on star/planet formation.  ESPaDOnS and Narval, the twin high-resolution 
spectropolarimeters respectively mounted on CFHT \citep{Donati03,Donati06d} and on the 2m T\'elescope Bernard Lyot (TBL), already allowed to unveil 
for the first time magnetic topologies of PMS objects \citep{Donati05, Donati10, Skelly10, Donati13, Donati14} and to detect the youngest 
known hot Jupiters (hJs) to date \citep[][see Fig.~\ref{fig:lsfp}]{Donati16, Donati17, Yu17}, demonstrating that planet formation and 
planet-disc interaction are both quite efficient on timescales of less than 2~Myr.  However, our knowledge of magnetic fields and 
planetary systems of PMS stars is still fragmentary, the intrinsic faintness of these objects in the visible drastically limiting their 
accessibility even to the most sensitive instruments.  

SPIRou was designed to address these two forefront issues with unprecedented efficiency \citep{Delfosse13, Artigau14, Moutou15}.  By 
operating in the nIR (including the K band), it will offer maximum sensitivity to both M dwarfs and PMS stars.  Moreover, by coupling 
spectropolarimetry with high-precision velocimetry, SPIRou will allow us to model magnetic activity and to filter-out RV curves more 
accurately than previously possible, and thus to achieve major progress in our exploration of planetary systems of nearby M dwarfs and 
in our understanding of planetary formation at early stages of evolution.  
In particular, thanks to its widest nIR coverage among planet hunters coupled to its unique polarimetric and activity-filtering capabilities,  
SPIRou will be especially efficient at detecting and characterizing planets around late-M dwarfs whose high levels of 
magnetic activity are notorious.  

We describe below in more detail the science programmes 
underlying these two prime goals, to which the SPIRou Legacy Survey (SLS) of about 500~CFHT nights will be dedicated, and mention 
the many other exciting programmes that SPIRou will be able to efficiently tackle thanks to its unique observational assets.  
We also provide a technical description of SPIRou, outline the expected performances on which our ambitious Legacy Survey relies, 
and summarize the overall project characteristics in terms of schedule, budget and manpower.  
We finally conclude with SPIRou-related prospects over the next decade.

\section{Science with SPIRou}
\label{sec:science}

We detail below the two main science goals to which our SPIRou Legacy Survey is dedicated;  we also mention a few additional programmes 
that SPIRou will be able to tackle, and the worldwide science consortium thanks to which SPIRou is coming to life.  

\subsection{Planetary systems of nearby M dwarfs}

Much interest has recently been focused on planets of M dwarfs \citep{Bonfils13, Muirhead15}, with the conclusion that these stars host 
low-mass planets more frequently than Sun-like stars do \citep{Dressing15, Gaidos16}.  The recent discovery of a HZ planet around Proxima~Cen 
\citep[][see also dedicated section in this book]{Anglada16} further triggered the motivation to detect and study low-mass planets and 
planetary systems around nearby red dwarfs.  The main goals are to reveal the planet occurrence frequencies and system architectures, 
to investigate how they depend on the masses of the host stars (and thus on the masses and properties of the parent protoplanetary disc), 
and ultimately to better characterize the formation mechanism(s) that led to the observed distributions of planets and systems.  

Up to now, only a few such planets have been 
detected and characterized with RV observations, which required in particular focusing on the brightest M dwarfs as the only accessible 
targets for the few existing optical velocimeters capable of reaching a RV precision of 1~\ms.  This is clearly insufficient to achieve 
a proper statistical study of rocky exoplanets and more generally of exoplanetary systems around M dwarfs.  This constraint also drastically 
limits our chances of detecting transiting rocky planets in the HZs of the nearest stars, i.e., the only ones for which atmospheric 
characterization with the JWST will be possible \citep{Berta15}.

Carrying out an exploration of nearby M dwarfs extensive enough to detect and characterize hundreds of low-mass planets and planetary 
systems, whose existence is known, mandatorily requires RV observations in the nIR domain, where these stars are brightest.  This is what 
SPIRou aims at with the SLS, concentrating the effort in two main directions, (i) a systematic RV monitoring of a large sample of nearby 
M dwarfs (called the SLS Planet Search) and (ii) a RV follow-up of the most interesting transiting planet candidates to be uncovered by 
future photometric surveys (called the SLS Transit Follow-up).  In both cases, SPIRou will be observing in spectropolarimetric mode to 
simultaneously monitor stellar activity, unambiguously identify the rotation period (with which activity is modulated) and reconstruct the 
parent large-scale magnetic field triggering the activity.  This will enable to implement novel and efficient ways of filtering out the 
polluting effect of activity from RV curves \citep{Hebrard16}, and thus to boost the sensitivity of SPIRou to low-mass planets.  
This option will turn especially useful for late-M dwarfs, many of which are rather active as a result of their higher rotation rates 
\citep{Newton16} and show RV activity jitters of several \ms\ \citep{Gomes12, Hebrard16}.  

The immediate objective of the SLS Planet Search is to: 
\begin{itemize}
\item identify at least 200 exoplanets with orbital periods ranging from 1~d to 1~yr around stars with masses spanning 0.08--0.5~\msun\ 
to derive accurate planet statistics as a function of stellar mass;
\item identify a few tens of HZ terrestrial planets orbiting nearby M dwarfs, thanks to which we will infer a better description of 
the different types of planets located in HZs; 
\item identifying several tens of multi-planet systems for studying the architecture of exoplanetary systems and their dynamical evolution;
\item identifying a large population of close-in planets to investigate how they form and interact with the magnetospheres of their 
host stars.
\end{itemize}

\begin{figure}[t]
\includegraphics[scale=1.35]{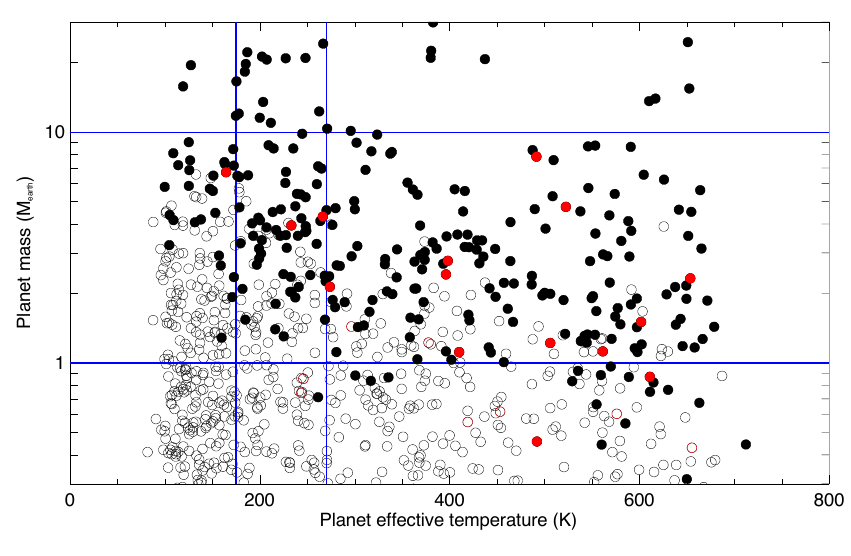}
\caption{Simulation of an example SLS Planet Search assuming a sample size of 360~M dwarfs and 55~visits per star, for a total survey 
time of 300~CFHT nights.  Filled circled indicate detected planets, open circles undetected ones and red circles (both filled and open) 
represent transiting planets. Blue lines show notional limits for the HZ, both in mass and temperature. Most planets more massive than 
2~\mearth\ located in the HZ are detected, including two transiting.  A sample of sub-\mearth\ planets hotter than 350~K is also 
detected.  Using a smaller sample with more visits (as we now propose) improves the detectability of multi-planet systems.  }
\label{fig:survey}
\end{figure}

Practically speaking, this implies carrying out a deep survey of at least 200~M dwarfs of different masses, with typically 100 visits 
per star (each yielding a spectrum with high-enough S/N to achieve 1~\ms\ RV precision).  Monte Carlo simulations (see, e.g., 
Fig.~\ref{fig:survey}) suggest that the SLS 
planet search should detect at least 200 new exoplanets, including 150 with masses $<$5~\mearth\ and 20 located in the HZs 
of their host stars.  The SLS should thus offer a yet unparalleled opportunity to explore the diversity of HZ Earth-like planets and 
planetary systems, and to reveal which of them are most common.  With improved statistics on planets / systems around M dwarfs, our 
survey will constrain models of planetary formation, in particular regarding the sensitivity of planet formation to initial 
conditions in protoplanetary discs and to the mass of the host stars.  The SLS will also probe the occurrence rate of sub-Earth-mass 
planets with orbital periods shorter than 10~d;  the discovery of 3 Mars-sized planets in close orbits around a M5 dwarf within the 
very small sample of late-M dwarfs in the Kepler field indicates that such worlds are likely to be very common \citep{Muirhead12} 
in addition to being potentially fruitful targets to characterize, their very tight orbits implying high probabilities of transit.  
The goal is to invest 275~CFHT nights in the SLS Planet Search.  

Regarding the SLS Transit Follow-up, the obvious goal is to use transiting planets as sensitive probes of their internal structures 
and atmospheres.  Ground-based nIR spectroscopy is indeed essential to this quest, spectroscopy being mandatory to validate the 
planetary nature of the transiting candidate planets detected around M dwarfs through photometric monitoring, but also to measure 
their masses from their host stars' RV-curve amplitudes.  Whereas Kepler showed that planets smaller than 4~\rearth\ are quite 
common, even in compact and coplanar multi-planet systems, it however failed to discover Earth twins close enough for further 
atmospheric characterization, or even to be validated and characterized through a RV follow-up.  The goal of future photometric 
surveys is thus to detect planet candidates around brighter stars, with a specific emphasis on nearby M dwarfs.

Among them, TESS, to be launched in 2018 and predicted to detect about 300 super-Earths, is the most promising.  By surveying the 
entire sky (with individual regions scrutinized on timescales of 27~d) and focussing on the brightest targets, TESS will monitor the 
light curves of more than 200,000 targets at a high temporal cadence, and 100$\times$ more at a slow cadence, during a nominal 2-yr 
mission. From Kepler statistics, TESS predicts the discovery of several hundreds of Earth-like and super-Earth planets, in addition to 
thousands of icy and gas giants, for stars brighter than I=12 \citep{Sullivan15}.  Most super-Earth candidates that TESS will detect 
will orbit around M dwarfs, with less than 30\% accessible to optical velocimeters.  SPIRou will thus play an essential role in validating 
planet candidates and in measuring their masses \citep{Santerne13}.  

The goal is to carry out a RV follow-up of the 50 most interesting transiting planet candidates uncovered by future photometric surveys 
including TESS, for a total of 100~CFHT nights.  Monte Carlo simulations indicate that SPIRou has the capacity to validate and 
characterize Earth-mass planets orbiting mid-M dwarfs, including those located in the HZs of their host stars on which SPIRou will 
concentrate (see Fig.~\ref{fig:survey}).  

As SPIRou will simultaneously secure spectropolarimetric and velocimetric data, it will allow us to model at the same time the 
large-scale magnetic topologies and the associated activity of the host stars, and thus to filter out the RV jitter from RV curves 
and enhance the sensitivity to low-mass planets \citep{Hebrard16}.  Investigating how large-scale fields of M-dwarfs vary with 
stellar parameters \citep{Morin08b, Morin10} will unveil how dynamo processes behave in fully-convective bodies, and how such dynamo 
fields can either degrade or improve habitability depending on whether they anchor in the host stars or in their planets \citep{Vidotto13, Gudel14}.  

By providing a wide and homogeneous set of nIR spectra for all types of M dwarfs, the SLS exoplanet programmes will also give the 
opportunity to assess theoretical atmospheric models of cool and very cool stars in much more details than what is currently 
possible;  in particular, it will make it possible to further constrain key physical processes occurring in the atmospheres of 
very-cool stars and affecting their thermal and convection patterns \citep{Rajpurohit13, Allard13, Passegger16}.  

\subsection{Magnetic fields and star/planet formation}

Studying how Sun-like stars and their planetary systems form comes as an obvious complement to the direct observation of exoplanets 
in our quest to understand the emergence of life.  The second main goal of SPIRou is thus to explore the impact of magnetic fields 
on star/planet formation, by detecting and characterizing magnetic fields of low-mass PMS stars and their inner accretion discs.  

By controlling accretion, triggering outflows and jets, and producing intense X-rays, magnetic fields critically impact the physics 
of PMS stars \citep{Baraffe10, Feiden16} and of their accretion discs \citep{Shu07}, and largely dictate their angular 
momentum evolution \citep{Bouvier07}.  In particular, magnetic fields are thought to couple accreting PMS stars with their discs;  
more specifically, fields carve magnetospheric gaps in the central disc regions and trigger funneled inflows \& outflows from the inner 
discs, forcing the host stars to spin down \citep{Romanova04, Romanova11, Zanni13, Davies14}.  Magnetic fields presumably affect 
planet formation as well \citep{Johansen09}, can stop or even reverse planet migration \citep{Baruteau14}, and may prevent close-in 
planets, including hJs \citep{Lin96, Romanova06}, from falling into their host stars.  

\begin{figure}[t]
\includegraphics[scale=0.74]{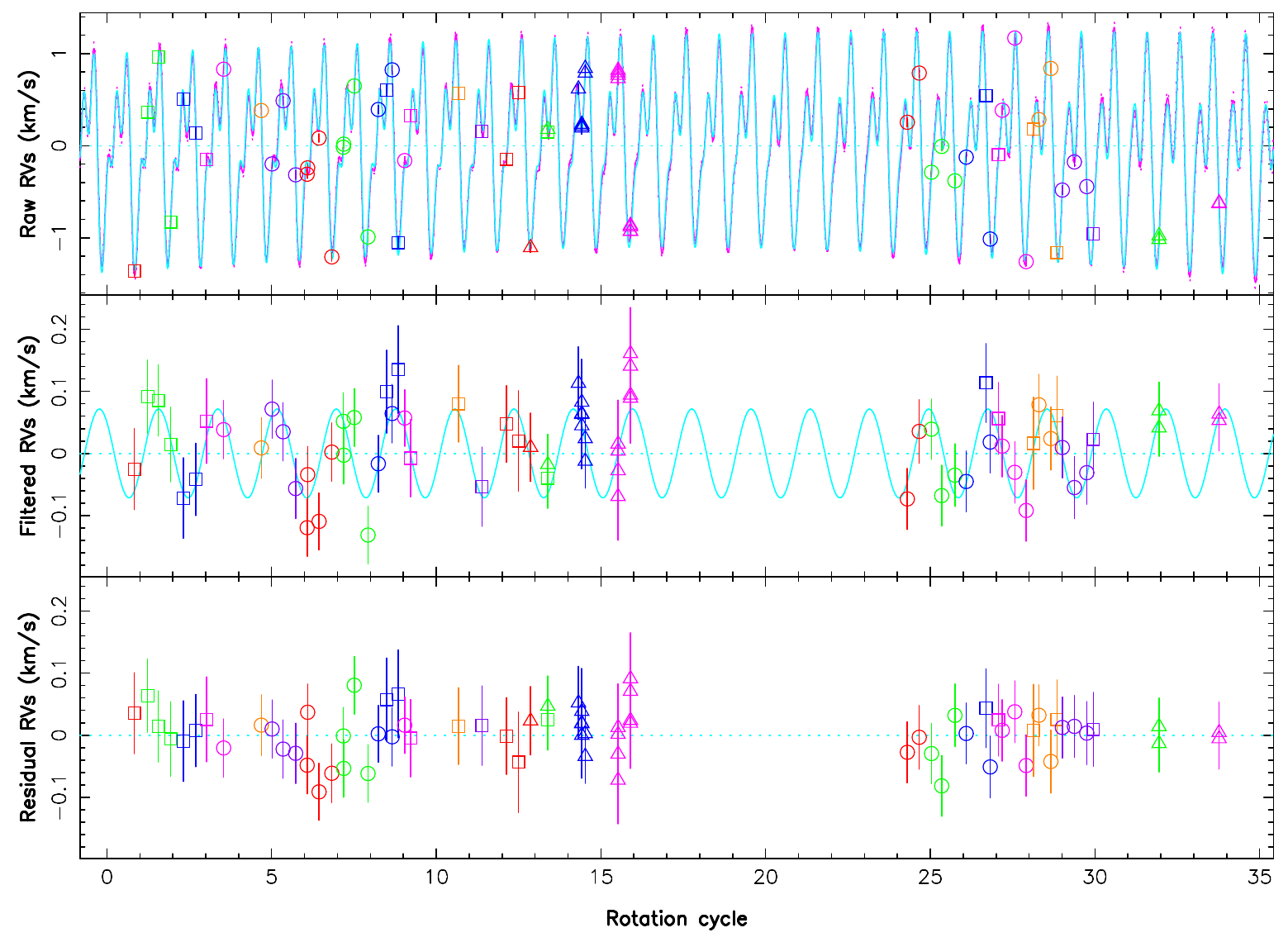}
\caption{Modelling the RV variations of the wTTSs V830~Tau with a Gaussian Process for the activity jitter and a 0.7~Jupiter mass 
planet in a close-in circular orbit.  Raw, filtered and residual RVs are shown in the top, middle and bottom panels respectively.  
Observations are plotted as open symbols;  the activity jitter and the planet RV signal are depicted as cyan lines in the top and 
middle panels respectively, whereas the combination of both is shown as a purple line in the top panel.  The modelling unambiguously 
reveals the existence of a hJ despite its RV signal being 15$\times$ weaker than the activity jitter.  The rms dispersion of the 
RV residuals is 37~\ms, i.e., close to the instrumental RV precision \citep{Donati17}. } 
\label{fig:v830tau}
\end{figure}

The initial exploration carried out with the optical spectropolarimeters ESPaDOnS at CFHT and Narval at TBL revealed that TTSs host strong large-scale 
magnetic fields \citep{Skelly10, Donati13, Donati14} of dynamo origin \citep{Donati12} and whose topologies largely reflect the 
internal structures or the host stars \citep{Gregory12}.  Intense fields were also unambiguously detected in the inner regions of 
an outbursting accretion disc \citep{Donati05}.  More recently, newborn close-in giant planets, including the youngest hJ known to 
date \citep[][see, e.g., Fig.~\ref{fig:v830tau}]{Donati16, Donati17}, were discovered around PMS stars \citep{David16, Yu17}, providing new 
evidence that planet-disc interactions 
plays a key role in planet formation and are likely to shape the early architecture of planetary systems.  Despite this progress, 
our knowledge of magnetic topologies of low-mass PMS stars, and of their planet-forming accretion discs whenever relevant, is still 
fragmentary, and very few observational constraints are available to test models of star/planet formation.  

Extending spectropolarimetric observations to the nIR domain is the most logical step forward.  Low-mass PMS stars are indeed brighter 
in the nIR than in the optical, especially embedded class-I protostars, whereas the Zeeman effect is stronger at longer wavelengths.  
SPIRou thus has the potential to explore much larger samples of PMS stars than previously possible, and to vastly improve the statistical 
significance of our current results, in particular for embedded protostars and accretion discs for which very little information is available.  
To achieve this, we need to initiate a large survey aimed at thoroughly exploring how magnetic fields impact star/planet formation.  
This is the goal of the third SLS component, that we call the SLS Magnetic PMS star/planet survey.   

In this aim, SPIRou will monitor 20~low-mass class-I protostars and 40~cTTSs with their accretion discs, as well as 80~wTTSs, selected 
in the 3 closest star forming regions (Tau/Aur, TW~Hya and $\rho$~Oph/Lupus).  As the missing link between the youngest class-0 
protostars whose fields are surveyed with ALMA / NOEMA at mm wavelengths \citep{Maury10} and the older cTTSs observed with optical 
instruments like ESPaDOnS, class-I protostars are key for fingerprinting the impact of magnetic fields on star/planet formation.  
Their reputedly strong fields \citep{Johns09}, whose large-scale topologies are still unknown, will tell us how dynamos \citep{Gregory12} 
and magnetospheric accretion \citep{Romanova08, Zanni13, Blinova16} behave when accretion is stronger and more episodic than in cTTSs, 
how large a magnetospheric gap these fields carve at disc centre, and how stars react to this process \citep{Baraffe10, Feiden16}.  
The 40~cTTSs will be medium to strong accretors or very-low-mass stars for which very few spectropolarimetric observations exist so 
far, thanks to which we will sample the whole range of masses and accretion patterns \citep{Cody14, Sousa16}.  In both cases, SPIRou will 
have a chance to detect and characterize the magnetic fields of their inner discs, and to pin down at the same time the main disc 
properties \citep{Carmona13}.  

By monitoring a sample of 80~wTTSs and building on the discovery of the youngest known hJs \citep{Donati16, Donati17, Yu17}, SPIRou will 
characterize the population of newborn close-in giant planets at early evolutionary stages, estimate their occurrence frequency and 
compare it with that of mature Sun-like stars, thereby implementing a novel way of fingerprinting planet formation and planet/disc 
interactions \citep{Baruteau14}.   
This survey will also logically complement that of accreting PMS stars;  SPIRou will characterize the magnetic topologies of the 
observed wTTSs to complete the magnetic panorama of young Sun-like stars.  More specifically, SPIRou will work out how 
large-scale fields change with mass and age, at ages similar to those of cTTSs but with no impact from accretion, and infer how 
these fields affect the AM evolution of wTTSs through strong winds and massive prominences, and what role they play in 
star/planet interactions \citep{Strugarek15, Vidotto17}.  

\subsection{The SPIRou Legacy Survey (SLS) and the synergy with major observatories} 

Completing the SLS requires a total of 500~CFHT nights over a timescale of 5~years.  Whereas CFHT already agreed on investing 300~nights 
over 3~years for the SLS as a reward to the SPIRou consortium once SPIRou is installed on the telescope, the full SLS allocation of 
500~nights is still pending for an official confirmation once SPIRou performances are validated in the lab.  

By providing to the whole CFHT community a unique and homogeneous collection of nIR high-resolution spectra of M dwarfs and 
PMS stars that will be used for a wide range of various purposes, the SLS features an obvious Legacy dimension.  We plan to further 
enhance it through a world-wide accessible data base including additional material such as stellar fundamental parameters, precise 
RVs, Zeeman signatures and abundances of the main telluric molecules.  

All SLS programmes will exploit numerous synergies with major ground and space facilities like TESS and later-on PLATO for characterizing 
transiting planetary systems around M dwarfs, and the JWST for investigating planetary atmospheres.  ALMA will nicely complement SLS observations 
by scrutinizing star/planet formation and protoplanetary discs of PMS stars beyond a few au's from the host stars, whereas VLA/VLBA and 
LOFAR will be key for studying star/planet interactions in newborn hJs \citep{Bower16, Vidotto17}.  

\subsection{Additional science goals} 

SPIRou will also tackle a large number of additional science programmes beyond the two main ones forming the SLS.  

Studying weather patterns in the atmospheres of brown dwarfs is a particularly exciting option.  These objects are known to 
exhibit photometric variations on short timescales \citep{Artigau09}, attributed to the presence of atmospheric clouds rotating 
in and out of view, and subject to intrinsic variability.  Using tomographic imaging applied to time-series of high-resolution 
spectra, one can recover surface maps of the cloud patterns \citep{Crossfield14} and potentially their temporal evolution as 
well.  With its high sensitivity and large spectral domain, SPIRou will come as an ideal tool for carrying out such studies.  

SPIRou is equally well suited for investigating the dynamics and chemistry of planetary atmospheres in our Solar System \citep{Machado14,Machado17}, 
and potentially of giant close-in exoplanet atmospheres as well, even when not transiting \citep{Snellen10, Brogi12}.  Last but not 
least, SPIRou will also offer the opportunity of studying at high spectral resolution extremely metal-poor stars as relics of 
the early universe, providing us with precious clues about the chemical evolution and formation of the Milky Way \citep{Reggiani16}.  

\subsection{The SPIRou science consortium} 

The SPIRou science consortium gathers over 100 scientists from more than 30 research institutes in 11 different countries.  It 
includes in particular a strong French and Canadian core team, illustrating the fruitful collaboration built up over the last few 
decades of shared observing effort at CFHT.  The consortium also involves scientists from more recent CFHT partners such as Brazil 
and Taiwan, as well as a small number of experts from non-CFHT countries, involved in the construction of SPIRou (e.g., Switzerland, 
Portugal) or bringing critical expertise to the analysis of the SLS data (e.g., UK).

\section{The SPIRou spectropolarimeter / velocimeter}
\label{sec:instr}

\begin{figure}[t]
\includegraphics[scale=1.05]{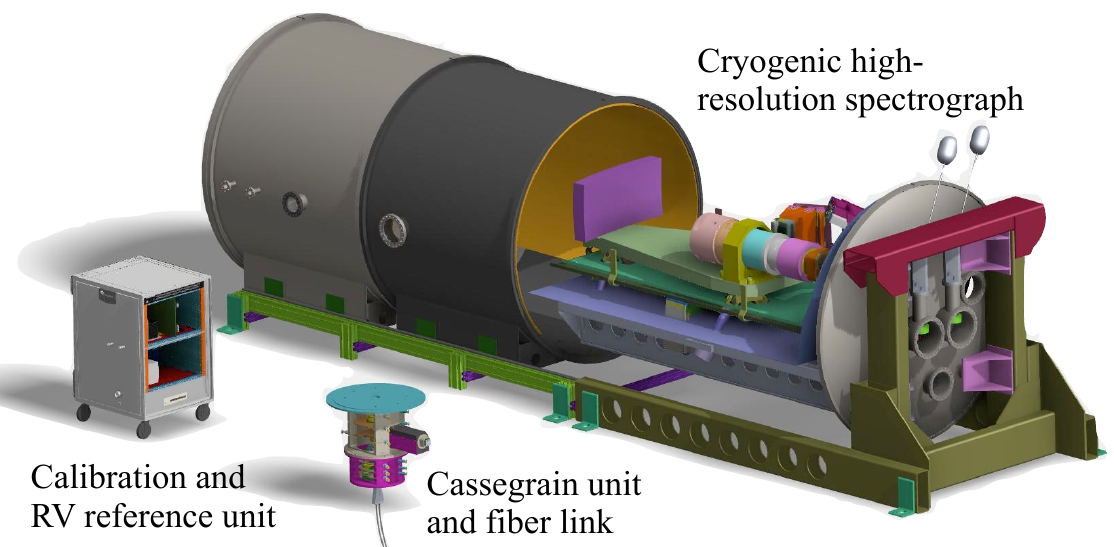}
\caption{CAD view of the main SPIRou units.} 
\label{fig:spirou}       
\end{figure}

SPIRou is a direct heritage from previous successful instruments, namely HARPS at the 3.6-m ESO telescope \citep{Pepe03} and ESPaDOnS 
at CFHT \citep{Donati03,Donati06d}, and whose overall characteristics are listed in Table~\ref{tab:spi}.  In particular, SPIRou includes a 
cryogenic high-resolution spectrograph inspired from the evacuated spectrograph of the HARPS velocimeter, a Cassegrain unit derived 
from the ESPaDOnS spectropolarimeter, a fiber-feed evolved from those of ESPaDOnS and HARPS, and a calibration/RV reference unit 
mirroring that of HARPS (see Fig.~\ref{fig:spirou}).  We describe these various technical units below in more details.  

\begin{table}
\caption[]{Main technical characteristics of SPIRou} 
\center{
\begin{tabular}{cc}
\hline
Spectral range in single exposure & 0.978--2.437~$\mu$m with no gaps \\
Radial-velocity stability         & better than 1~\ms   \\
Spectral resolving power          & $>$70\,000          \\
Detector array                    & H4RG-15 HgCdTe array, 4096$^2$ 15~$\mu$m pixels \\ 
Diffraction grating               & 306$\times$154~mm 22~gr/mm R2 grating from Richardson-Lab \\ 
Cross-dispersing prism train      & 2 ZnSe prism and 1 silica prism (size 190$\times$206~mm) \\ 
Velocity bin of detector pixel    & 2.3~\kms            \\ 
Throughput performances           & S/N=110 per 2.3~\kms\ bin at K$\simeq$11 in 1~hr for a M6 dwarf \\ 
Polarimetric performances         & circular \& linear, sensitivity 10~ppm, crosstalk $<$2\% \\ 
Spectrograph temperature          & 80~K, thermal stability 2~mK rms (goal 1~mK) on 24~hr \\ 
\hline
\end{tabular}}
\label{tab:spi}
\end{table}

\subsection{The Cassegrain unit and calibration tools} 

\begin{figure}[t]
\includegraphics[scale=0.6]{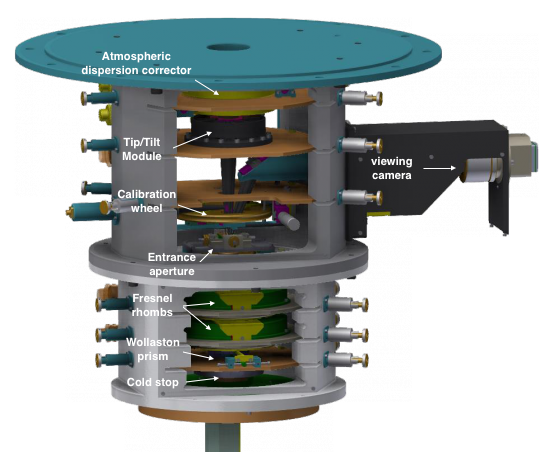}
\caption{CAD view of the middle and lower modules of the SPIRou Cassegrain unit.} 
\label{fig:cass}       
\end{figure}

The Cassegrain unit consists of 3 modules mounted on top of each other and fixed at the Cassegrain focus of the telescope.  
A CAD view of the lower two modules is shown in Fig.~\ref{fig:cass}.  

The upper Cassegrain module essentially serves as a mechanical interface with the telescope, ensuring that the instrument aperture 
is ideally placed with respect to the telescope focus.  It also includes an option for feeding the instrument with a fully polarized beam, 
allowing one to achieve a complete polarimetric diagnostic of all optical components above the polarimeter, including those located 
above the entrance aperture (see below).  

The middle Cassegrain module includes several facilities for either calibration or observation purposes.  It first includes 
an atmospheric dispersion corrector (ADC) cancelling atmospheric refraction in the incoming beam (down to a precision better 
than 0.03\arcsec\ up to airmasses of 2.5).  It also features a tip-tilt module (TTM) stabilizing the image of the star at the instrument 
aperture, enough to ensure that the entrance image averaged over the whole exposure (of at least 1~sec) is stable to better than 
0.05\arcsec\ rms.  This device works in conjunction with a SWIR viewing camera (sensitive to the J and H bands), looking at the instrument 
entrance aperture and sending back information to the TTM at a frequency of up to 50~Hz.  The main goal of the TTM is to minimise shifts  
of the stellar image with respect to the entrance aperture, and the systematic RV errors that may result from these shifts.  The final 
element of this module is a calibration wheel allowing one to inject light from the calibration unit (see below) into the instrument, 
and to polarize it linearly if need be.  

The lower Cassegrain module first includes a focal reducer turning the f/8 incoming beam entering the 1.3\arcsec\ circular instrument 
aperture, into a f/4 beam with which optical fibers are fed with minimum Focal Ratio Degradation (FRD).  This focal reducer is made of 
a doublet and a triplet working at infinite conjugate ratio, both optimized for the specific needs of SPIRou.  The entrance aperture 
is located at the centre of a tilted mirror reflecting back to the SWIR viewing camera the incoming light that does not enter the 
instrument.  This module also features an achromatic polarimeter located within the focal reducer, consisting of two 3/4-wave ZnSe dual 
Fresnel rhombs coupled to a Wollaston prism, splitting the incoming beam into 2 orthogonally polarised beams and feeding twin optical 
fibers.  By tuning the orientation of the rhombs, one can measure the amount of either circular or linear polarisation in the incoming 
stellar light.  By coating one of the internal reflection surfaces of each rhomb, we can ensure that the polarimetric analysis is 
achromatic to better than 0.5\degr.  Finally, this module includes a cold pupil stop located after the Wollaston prism, and blocking 
the thermal emission from the telescope to reduce the instrument thermal background in the reddest section of the spectral range.  
A more detailed account of the Cassegrain unit can be found in \citet{Pares12}.  

The Cassegrain unit can be fed from the calibration unit through the calibration wheel mentioned above.  This calibration unit 
provides light from the various lamps needed to calibrate observed spectra.  SPIRou uses a halogen lamp to collect flat field exposures 
(for tracking orders on the detector and correct for the spectral response of the instrument), a U/Ne hollow cathode for arc spectra 
(to obtain a very precise pixel to wavelength calibration), and an evacuated temperature-stabilised Fabry-Perot etalon featuring tens 
of thousands of sharp lines (to track the shape of the slit and monitor spectral drifts in the instrument).   The calibration unit can 
also directly feed light into the cryogenic spectrograph through the RV reference fiber (see below), thanks to which instrumental drifts 
can be corrected down to a precision of better than a few 0.1~\ms.  The calibration unit is described in more details in \citet{Boisse16}.  

\subsection{The fiber link and pupil slicer} 

The fiber link includes two science fibers conveying the light from the twin orthogonally polarized beams coming out of the Cassegrain 
polarimeter into the cryogenic spectrograph.  This item consists of a dual 35-m long 90-$\mu$m diameter circular fluoride fiber engineered 
by Le~Verre~Fluor\'e (LVF) from purified material, ensuring a throughput of at least 90\% over the entire spectral range of SPIRou.  The fiber 
link also includes a third shorter fiber, called the RV reference fiber, with which light from the calibration lamps can be fed directly 
into the spectrograph.  A triple hermetic feedthrough is used to inject light from the three fibers within the spectrograph.  Including 
transmission, FRD and connector losses, the fiber link provides an average transmission over the spectral range of $\simeq$70\%.  

\begin{figure}[t]
\hbox{\includegraphics[scale=0.95]{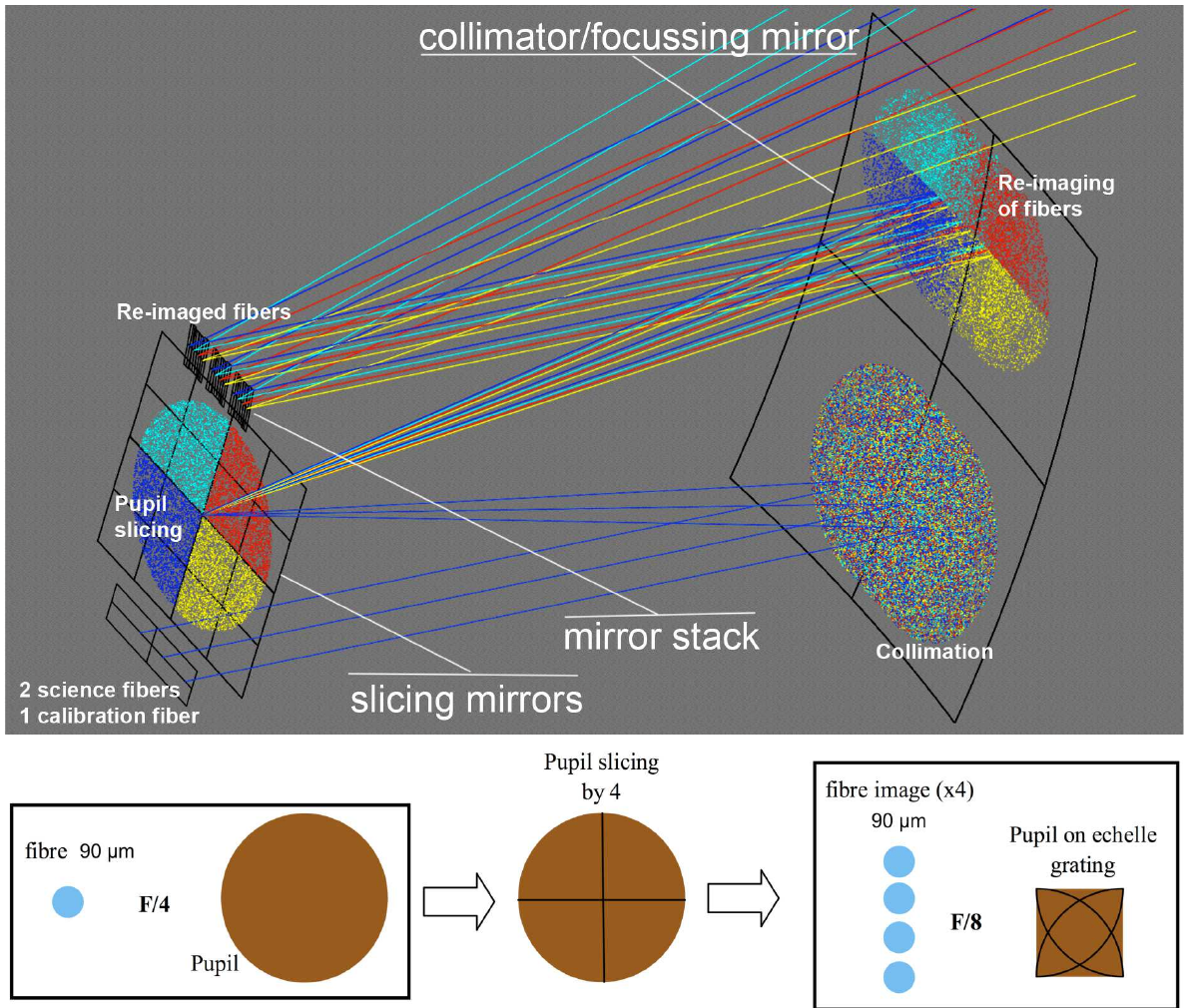} \vspace{1mm}} 
\includegraphics[scale=0.1015]{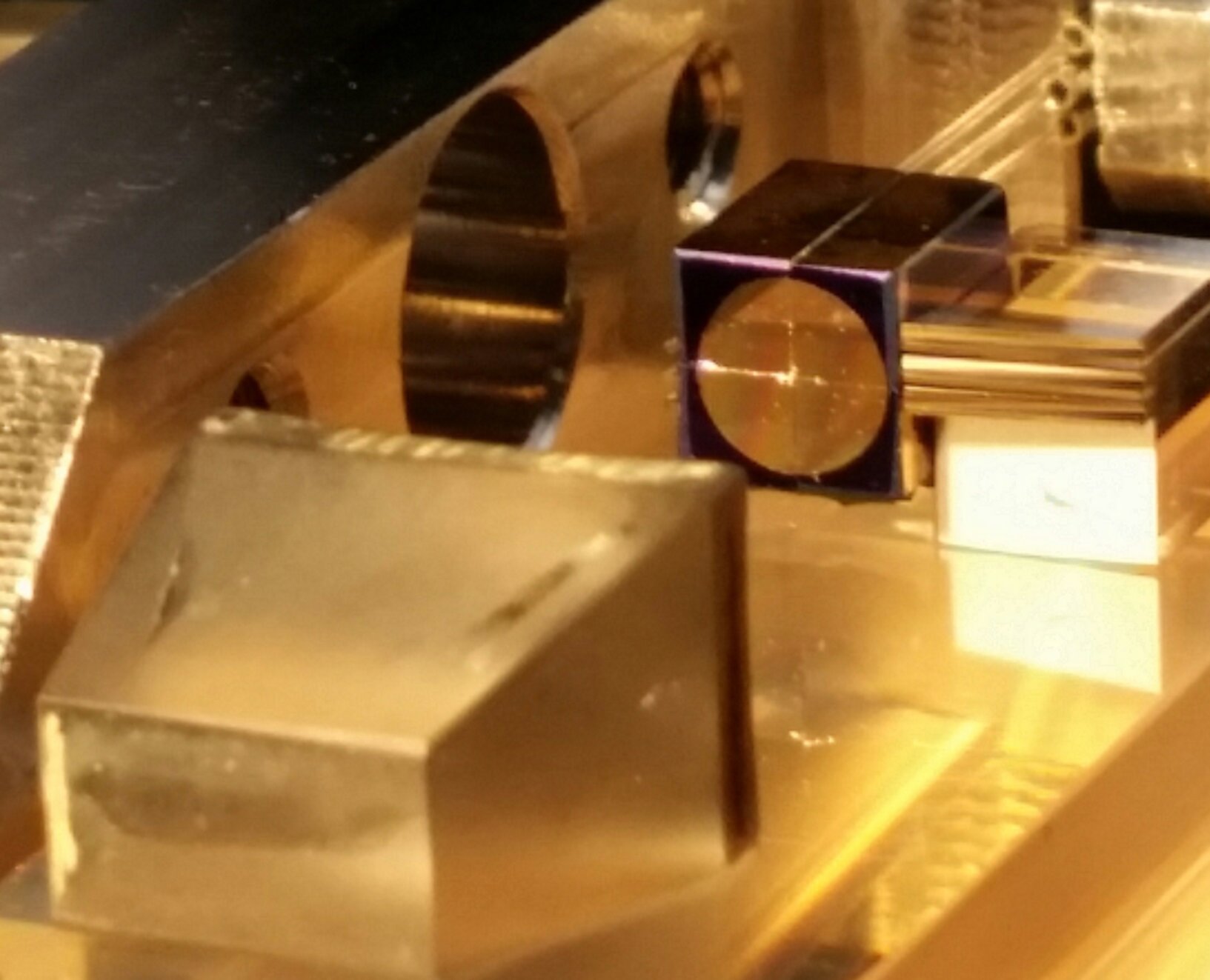}\includegraphics[scale=0.0482]{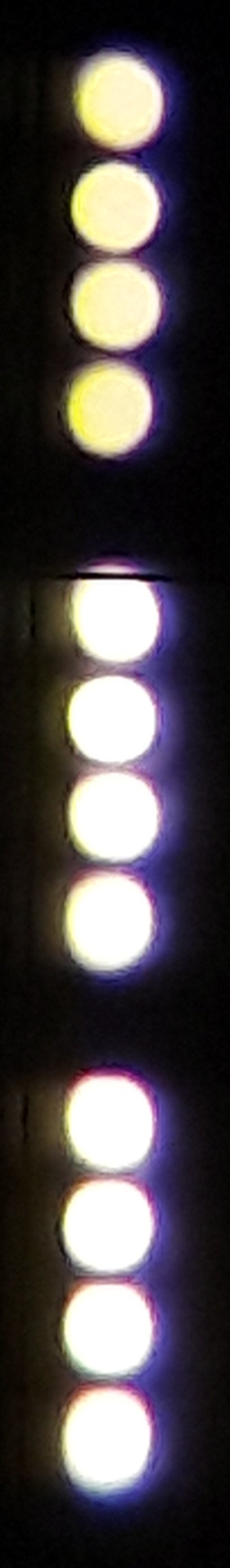}\includegraphics[scale=0.06]{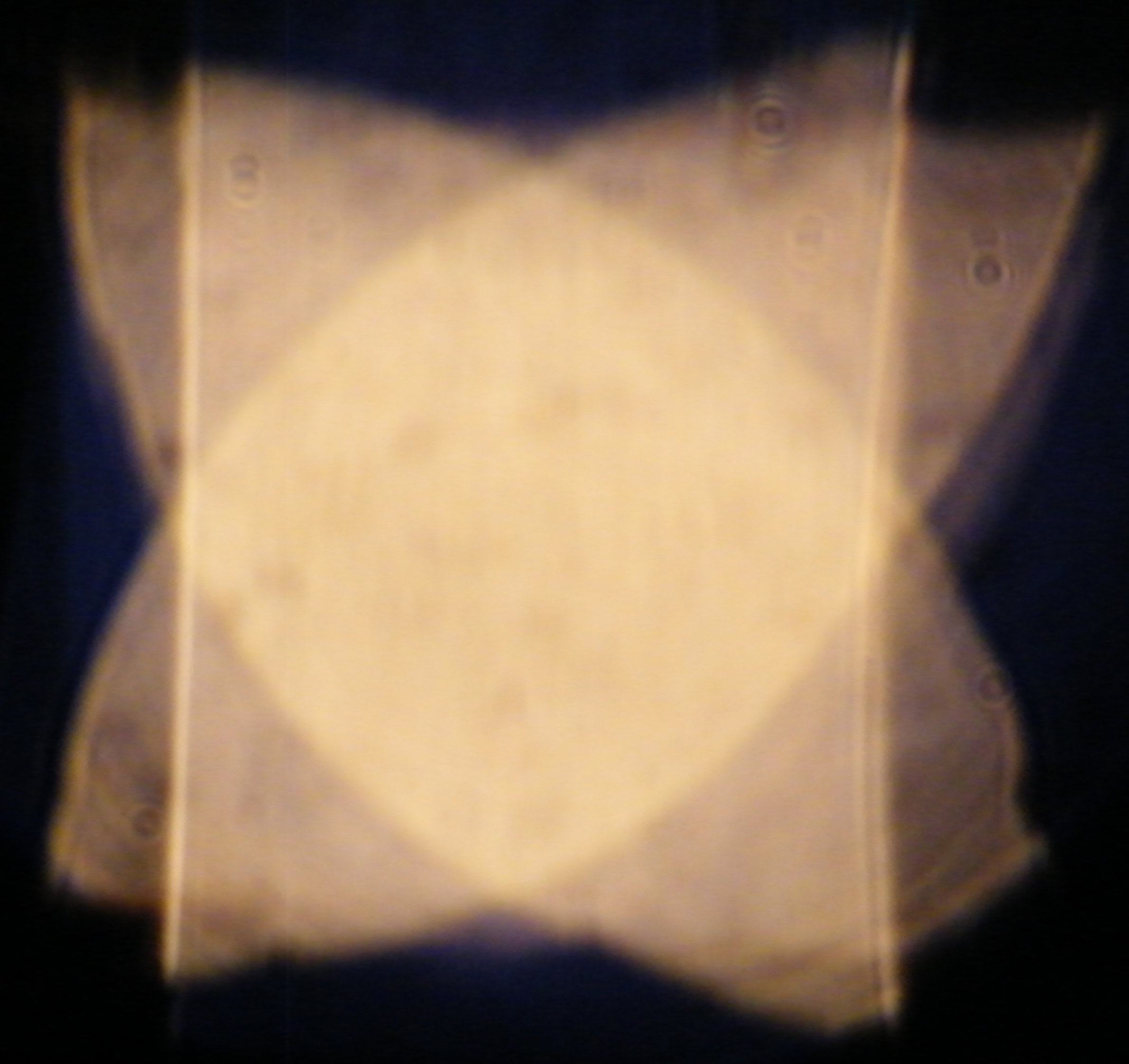}
\caption{{\bf Top panel:} Optical view of the SPIRou pupil slicer. {\bf Middle panel:} Schematic view of the pupil slicing principle 
for a single fibre.  {\bf Bottom panel:} Test pupil slicer assembled by WinLight (left), and corresponding slit / pupil 
profiles for a slightly oversized beam aperture (right). }
\label{fig:slicer}       
\end{figure}

This fiber link also includes a pupil-slicer at spectrograph entrance to minimize injection losses without sacrificing the spectrograph 
resolving power.  This pupil slicer is fed through three 1.4-m long segments of 90-$\mu$m-diameter octagonal fiber (also engineered by LVF) 
coming from the triple hermetic feedthrough.  The combination of the circular and octagonal fibers ensures a scrambling of the near-field 
image between polarimeter output and spectrograph input of at least 1000.

The pupil slicer per se consists of two mirrors, one main collimator and a pupil-slicing 
mirror located at the focus of the collimator and slicing the pupil into 4 equal 90\degr\ sectors;  the twelve individual images (4 images 
for each of the three fibers) formed after a second pass through the collimator are focused on a stack of twelve small mirrors ensuring 
that the pupils of all individual beams overlap into a square pupil once imaged onto the spectrograph grating (see Fig.~\ref{fig:slicer}). 
In addition of being extremely compact (a few cubic centimeters), this device has the advantage of simultaneously ensuring a high throughput 
($>$90\%), a high resolving power ($>$70,000) and sliced images with identical shapes and flux distributions (as opposed to more 
conventional image slicers for which slices often have different shapes), with the result of maximising image stability and thus RV precision.  

A more detailed account of the fiber link and pupil slicer can be found in \citet{Micheau12} and \citet{Micheau15}.  The first test pupil 
slicer for SPIRou, built by Winlight, was recently delivered to IRAP / OMP (see Fig.~\ref{fig:slicer}, bottom panel) and was shown to perform 
nominally.

\subsection{The cryogenic high-resolution spectrograph} 

\begin{figure}[t]
\includegraphics[scale=0.31]{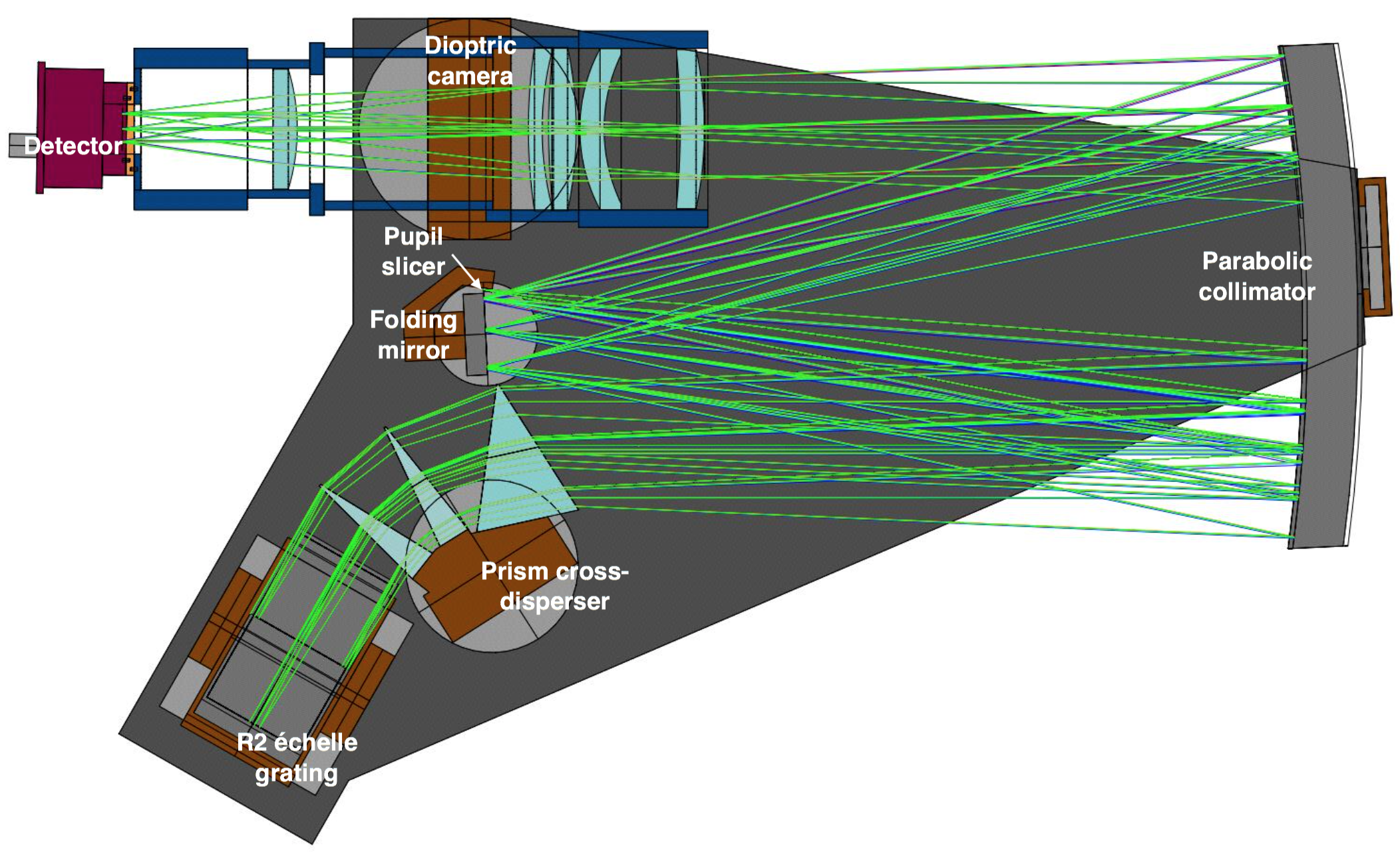}
\caption{Optical view of the SPIRou spectrograph.} 
\label{fig:spectro}       
\end{figure}

The bench-mounted high-resolution \'echelle spectrograph follows a dual-pupil design inspired from ESPaDOnS and HARPS (see Fig.~\ref{fig:spectro}).  
In the first half of the optical path, starting from the pupil slicer at the spectrograph entrance, the f/8 beam goes to the main collimator 
(1200~mm focal length, yielding a 150~mm square pupil) before being cross-dispersed with a double-pass triple-prism train (featuring two 
ZnSe prisms and an Infrasil one, both of height 206~mm) and dispersed in the perpendicular direction with a R2 \'echelle grating (of clear 
aperture $154\times306$~mm$^2$, w/ 23.2 gr/mm) from Richardson-Lab;  following a second pass on the collimator, the converging beam is 
reflected off the folding flat mirror and bounces back to the collimator.  In the second half of the optical path, the cross-dispersed 
\'echelle spectrum formed near the folding flat mirror is re-imaged onto a 4k$\times$4k H4RG detector (15~$\mu$m pixels) following a 
third pass on the collimator and a final focussing through a fully-dioptric 5-lens camera (500~mm focal length, clear aperture 220~mm).  

This design allows to record on the H4RG detector the entire spectral range of SPIRou (0.978--2.437~$\mu$m, 47 orders from \#78 to \#32) 
in a single exposure with no wavelength gaps between orders, at a spectral resolving power in excess of 70,000, and with an average velocity 
size of detector pixels of 2.28~\kms.  The spectrograph profile in the spectral direction is dominated by the slicer profile (equivalent full 
width at half maximum 4~\kms), with minor contributions from the detector pixels (1.8~\kms) and from the optical point-spread-function 
(1.5~\kms).  This design also ensures a high total throughput of 45\% detector included.  For a more complete description of the 
spectrograph optical design, the reader is referred to \citet{Thibault12}.  

The whole spectrograph is enclosed in a cryogenic dewar (of external diameter 1.73~m and length 2.87~m) and mounted on an optical bench 
supported at three points by an hexapod system from an internal warm support frame.  The spectrograph and optical bench are cooled down 
to 80~K and shielded by one active and three passive thermal screens, allowing one to stabilize the temperature of the bench and optics 
to within better than 2~mK.  This thermal stability ensures in particular that the spectral drift at detector level is $<$0.7~\ms\ on 
timescales of one night.  This drift can be monitored, and thus mostly corrected for, by recording the RV reference spectrum simultaneously 
with stellar spectra;  in this case, the residual spectral drift at detector level is reduced to 0.25~\ms.  Counting in all contributors 
to the RV error budget, SPIRou should achieve an RV precision of 1~\ms\ without the RV reference spectrum, and 0.75~\ms\ when using the 
simultaneous RV reference spectrum.  

The design of the cryogenic dewar and its thermal performances are described in more details in \citet{Reshetov12}.  The spectrograph 
cryomechanics and cooling system was demonstrated to be capable of ensuring a 80~K environment on the optical bench, with a thermal 
stability better than 2~mK rms on a timescale of 24~hr.  

\subsection{Controlling the instrument} 

The overall instrument control is fairly standard, the most challenging aspects being the control of the TTM on the mechanical side, and 
that of the dewar thermal stability on the temperature side.  More details can be found in \citet{Barrick12}.  

\subsection{The data simulator and reduction pipeline} 

\begin{figure}[t]
\includegraphics[scale=0.58]{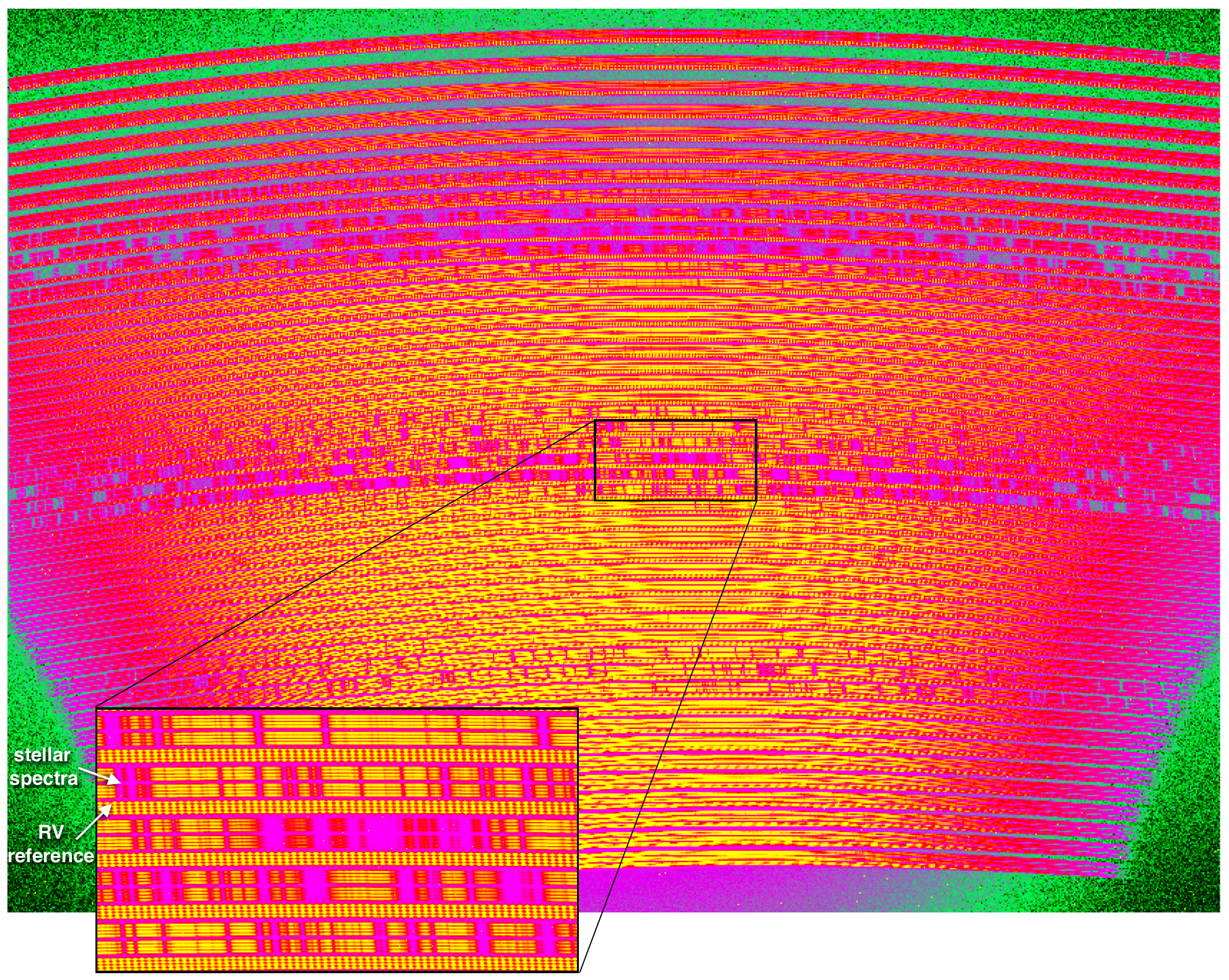}
\caption{Simulated raw SPIRou stellar frame, with a close up on the central part of orders \#53 to \#57.  The two stellar spectra of each 
order correspond to the orthogonal states of the selected polarisation, whereas the third spectrum tracks the Fabry-Perot RV reference.  
All three spectra feature four identical slices.  } 
\label{fig:simul}       
\end{figure}

A simulated raw SPIRou stellar frame is shown in Fig.~\ref{fig:simul} in a configuration where the RV reference is recorded simultaneously 
with the stellar spectra associated with the two orthogonal states of the selected polarisation.  Thanks to the tilted slit and the multiple 
slices and following the principles of optimal extraction \citep{Donati97b}, the SPIRou reduction pipeline is able to extract a spectrum with 
a velocity sampling of 1~\kms, i.e., 2.5$\times$ larger than the detector sampling of the original raw frame with no loss of information 
either in resolution or in signal to noise ratio.  This ensures in particular that extracted spectra are sampled on a wavelength grid fine 
enough for high-precision velocimetry.  More generally, tests on simulated data confirm that data reduction is compliant with all SPIRou 
science requirements.  

\subsection{Countdown to first light and science operation} 

\begin{figure}[t]
\includegraphics[scale=0.323]{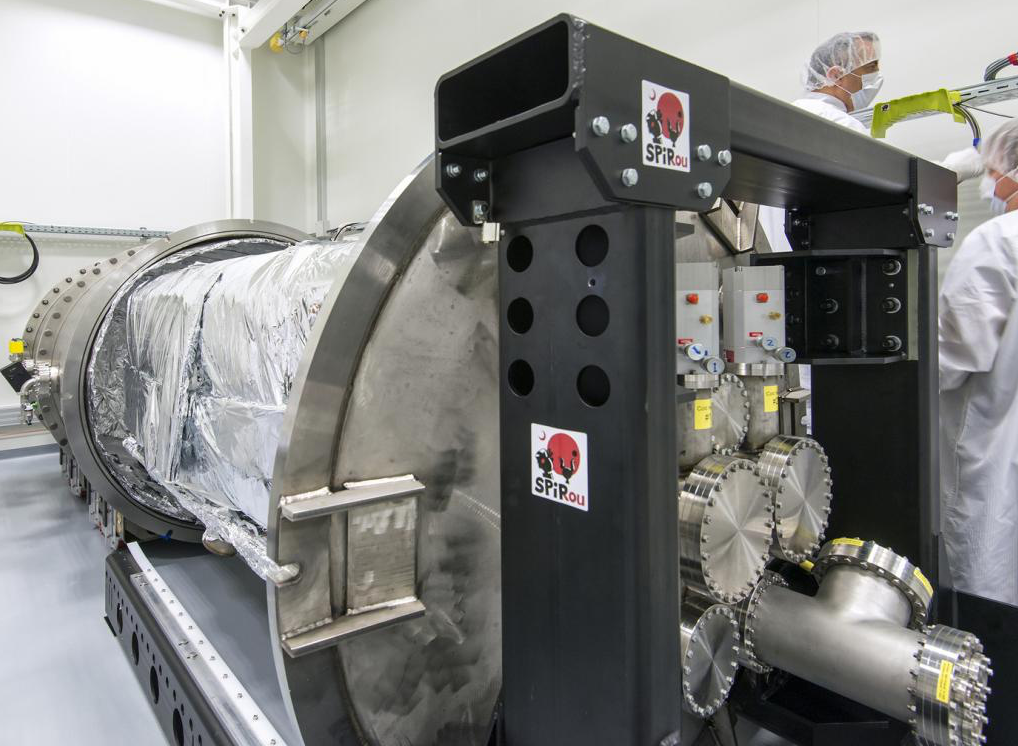}
\includegraphics[scale=0.223]{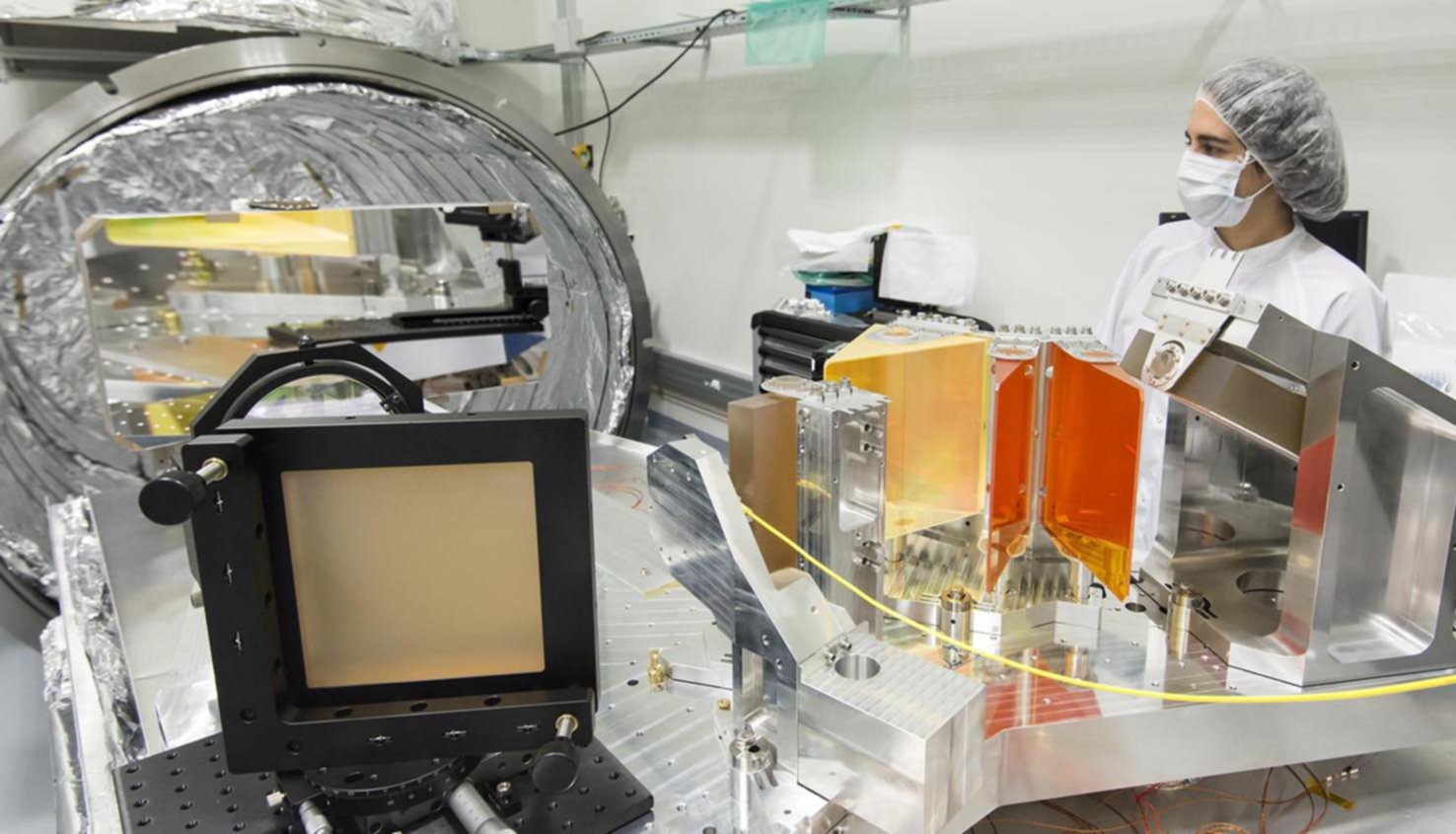}
\caption{{\bf Top panel:} Preparing for a SPIRou cryostat cool-down phase in the SPIRou clean room.  
{\bf Bottom panel:} Aligning the spectrograph optics on the bench while waiting for the spectrograph camera to reach IRAP / OMP;  
the grating, the prism train and the folding mirror are in the front, right-hand side, whereas the large parabolic collimator is 
in the rear, left-hand side (\copyright\ S~Chastanet, IRAP/OMP).} 
\label{fig:ait}
\end{figure}

\begin{figure}[t]
\includegraphics[scale=0.28]{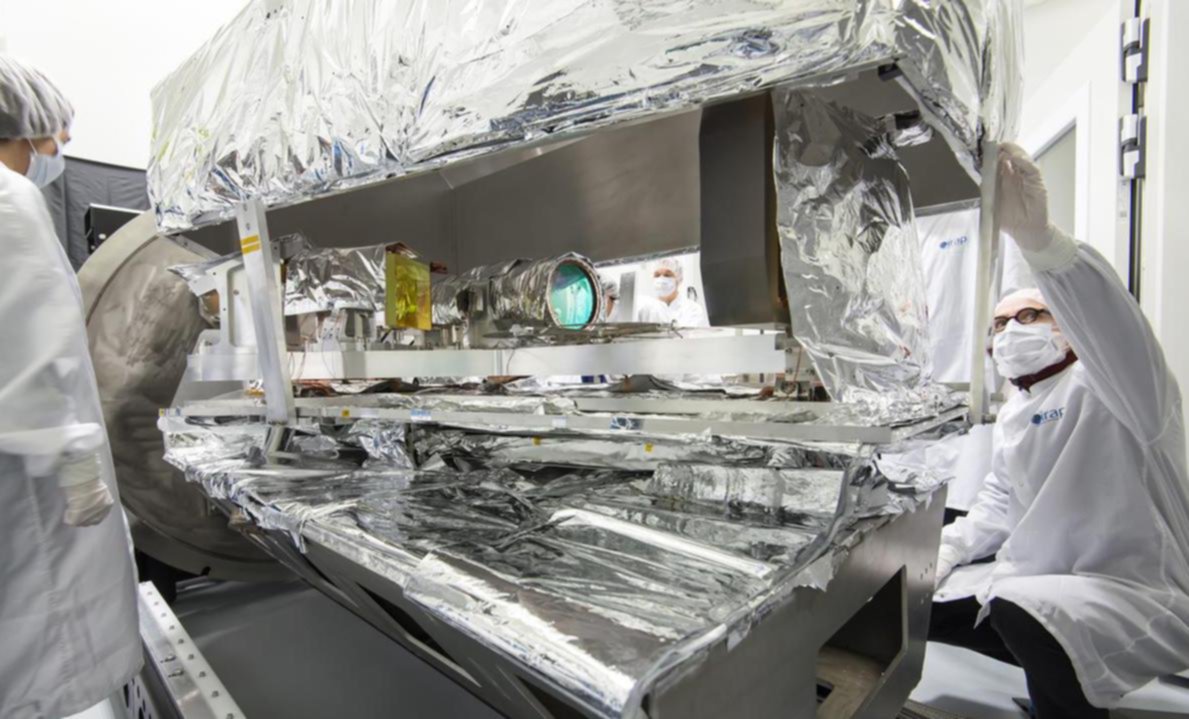}
\caption{Closing the SPIRou cryostat prior to the second cool-down cycle, with the spectrograph bench hosting
its full optical load (\copyright\ S~Chastanet, IRAP/OMP).}
\label{fig:ait2}
\end{figure}

\begin{figure}[t]
\includegraphics[scale=0.267]{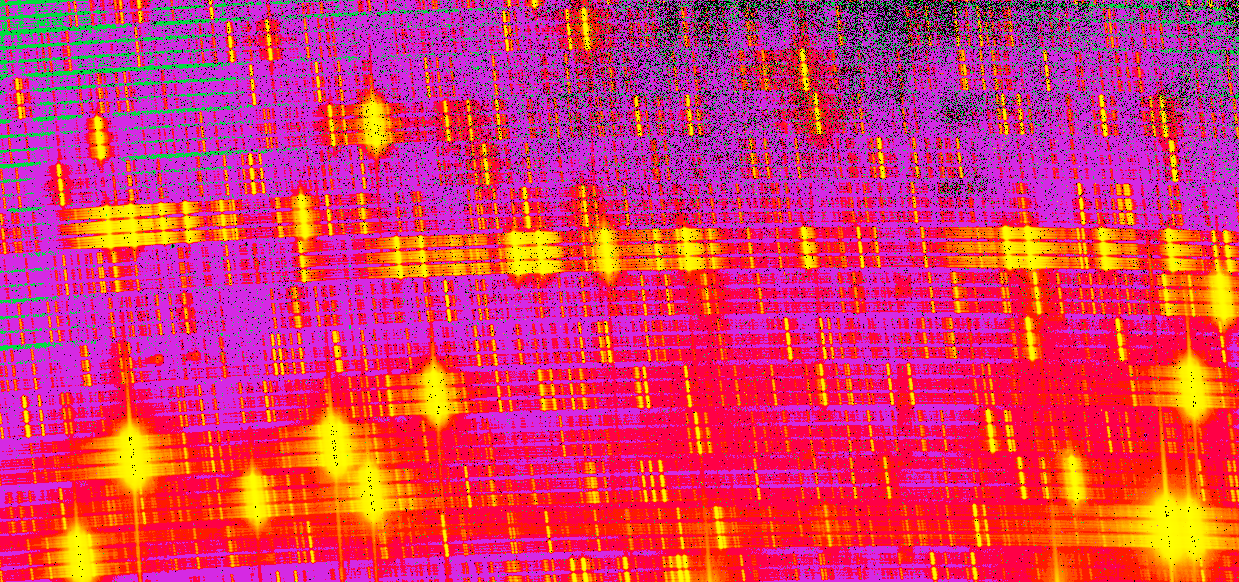} 
\vspace{-2.2mm}
\includegraphics[scale=0.267]{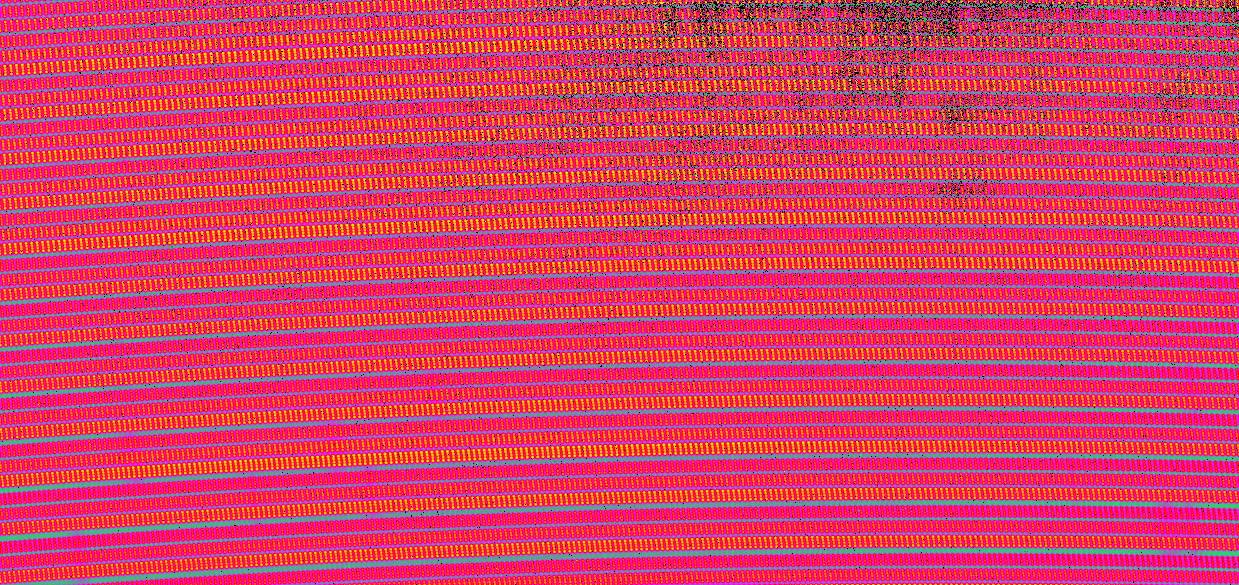} 
\caption{Blue orders of UNe hollow-cathode (top panel) and Fabry-Perot etalon (bottom panel) spectra recorded by SPIRou during 
the third thermal cycle in 2017 June, when the spectrograph detector was located only 50~$\mu$m away from the focal plane.  }  
\label{fig:spec}
\end{figure}

\begin{figure}[t]
\center{\includegraphics[scale=0.3,angle=270]{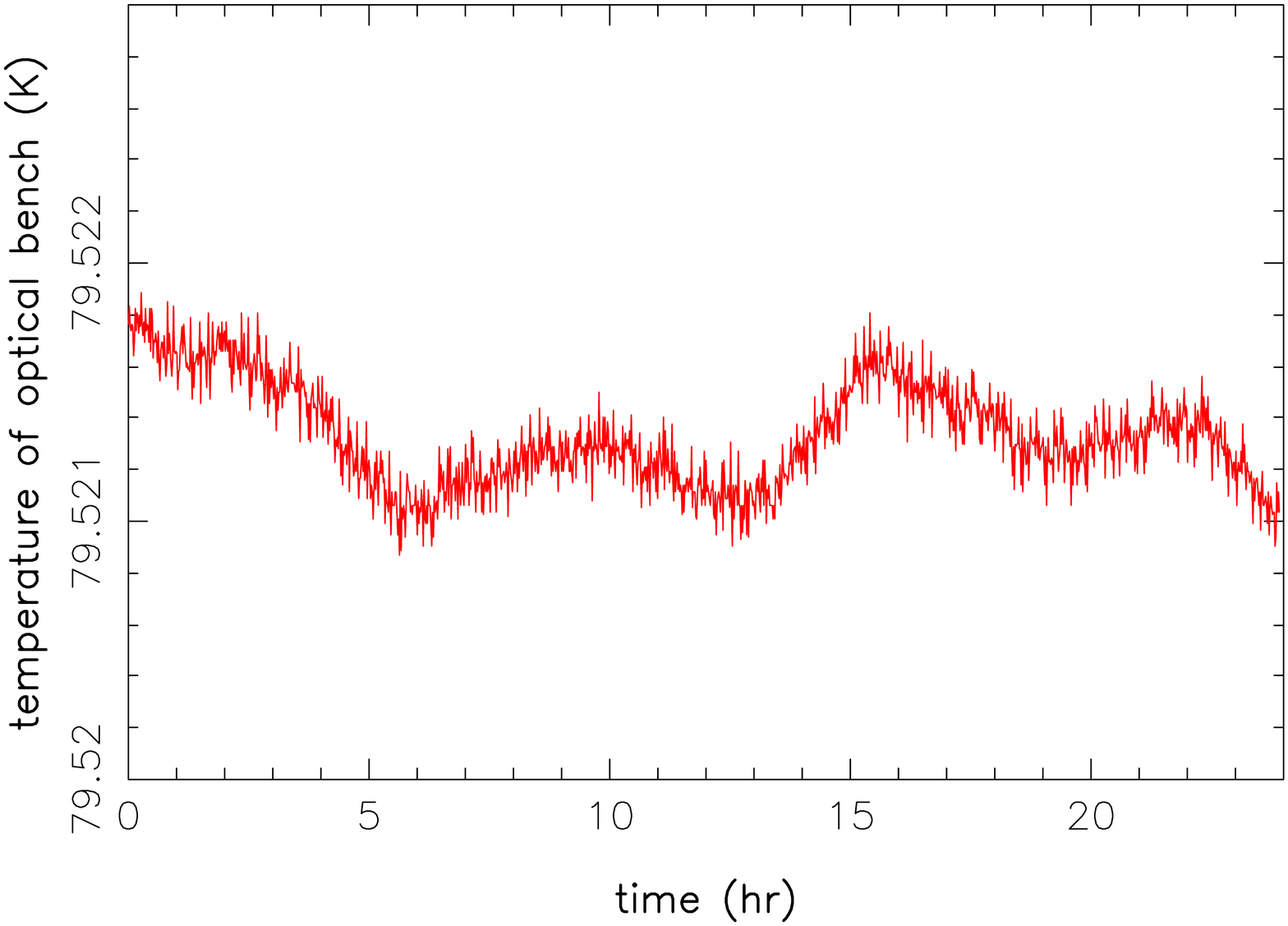}} 
\caption{Bench temperature at mid distance from the 3 control points on 2017 June 18, during the third SPIRou thermal cycle.  The rms 
thermal stability is found to be 0.2~mK, an order of magnitude better than our goal requirement of 2~mK.  } 
\label{fig:temp}
\end{figure}

SPIRou is currently being integrated in a clean room at IRAP/OMP in Toulouse, France (see Figs.~\ref{fig:ait} and \ref{fig:ait2}).  The spectrograph 
is now accurately aligned and almost in focus (see Fig.~\ref{fig:spec}).  The cryomechanics and its cooling system is found to perform optimally, with the thermal 
stability of the optical bench already proven to be significantly better than our drastic requirement of 2~mK rms over 24~hr (see Fig.~\ref{fig:temp}).  

Validation tests will be carried out until the final acceptance review, currently scheduled for 2017 September.  
Following packing and shipping, SPIRou will be installed at CFHT in 2017 November, 
with first light on the sky and technical commissioning planned for 2017 December onwards.  Finally, science verification will 
take place in 2018 Q2 before initiating SLS observations, hopefully by 2018 Q3.  Regular information on how integration progresses 
is available from the SPIRou website at URL {\tt spirou.irap.omp.eu}.  

\subsection{The SPIRou project team} 

The SPIRou project team gathers partners from 7 countries / corporations, and more specifically from: 
\begin{itemize}
\item France (IRAP/OMP in 
Toulouse in charge of the Cassegrain unit, the fiber link and slicer, the instrument integration and the overall project management, 
IPAG in Grenoble in charge of the optical design of the spectrograph, OHP/LAM in Marseille in charge of the calibration unit and of 
the reduction pipeline, plus technical inputs from LESIA in Paris); 
\item Canada (UdeM/UL in Montr\'eal and Qu\'ebec City, in charge of the spectrograph camera and detector, and NRC-H in Victoria, 
in charge of the spectrograph cryomechanics); 
\item CFHT (in charge of the overall instrument control); 
\item Taiwan (ASIAA in Taipei, in charge of the viewing camera and control of the TTM);  
\item Brazil (LNA in Itajuba, participating to the tests of the SPIRou optical fibers); 
\item Switzerland (Geneva Observatory, in charge of the RV reference module);  
\item Portugal (CAUP in Porto, participating to the construction of mechanical equipments for the instrument integration).  
\end{itemize}

All partners are also participating to the funding of SPIRou (at respective levels of 2.0~MEuros for France, 0.12~MEuros for Brazil, 
0.10~Meuros for Switzerland and Portugal, 0.07~MEuros for Taiwan, US\$ 2.0~M for CFHT and 0.7~M for Canada) for a total construction 
cost of 4.8~MEuros (assuming a change rate of 0.9~Euro per US\$).  The work effort associated with the construction amounts to a total 
of 60~FTEs (35 for France, 15 for Canada, 3 for CFHT and Switzerland, 2 for Taiwan, 1 for Brazil and Portugal), implying a consolidated 
cost for the whole instrument of about 10~MEuros.  

\section{Conclusions and prospects} 
\label{sec:concl}

SPIRou is a new-generation nIR spectropolarimeter / velocimeter for CFHT, currently in integration at IRAP/OMP, France.  
SPIRou covers a wide spectral nIR domain in a single exposure (0.98-2.44~$\mu$m) at a resolving power of 70~K, yielding 
unpolarized and polarized spectra of low-mass stars with a 15\% average throughput at a radial velocity (RV) precision of 1~\ms.  
It consists of a Cassegrain unit mounted at the Cassegrain focus of CFHT and featuring an achromatic polarimeter, coupled via a 
fluoride fiber link to a cryogenic spectrograph cooled down at 80~K and thermally stable at 2~mK rms on a timescale of 24~hr.  
Following integration and validation, SPIRou will be mounted at CFHT by 2017 Q4 for a first light scheduled in late 2017; 
science operation is expected to start in 2018 S2. 

SPIRou will focus on two main science topics, (i) the quest for habitable Earth-like planets around nearby M stars, 
and (ii) the study of low-mass star/planet formation in the presence of magnetic fields.  The SPIRou Legacy Survey is 
planning to dedicate about 500 nights over 5 years for carrying out forefront programmes on these two topics, in conjunction 
with other major ground and space facilities such as TESS, the JWST, ALMA and later-on PLATO and the ELT.  
SPIRou will also efficiently tackle many more programmes beyond these two main goals, from weather patterns on brown dwarfs 
to Solar-System planet and exoplanet atmospheres.  

In particular, we expect SPIRou to detect at least 200 new exoplanets around nearby M dwarfs, including 150 with masses smaller 
than 5~\mearth\ and 20 located in the HZs of their host stars.  SPIRou also plans to carry out a RV follow-up of the 50 most 
interesting transiting planet candidates to be uncovered by future photometry surveys like TESS.  Last but not least, SPIRou 
will achieve a thorough magnetic exploration of low-mass PMS stars to study how magnetic fields affect star/planet 
formation, and to assess how frequent hot Jupiters are at early stages of planetary formation and how critically they impact 
planetary system architectures.  

A twin version of SPIRou for TBL at Pic du Midi, nicknamed SPIP (for SPIRou-Pyr\'en\'ees), is already funded by R\'egion 
Occitanie / Pyr\'en\'ees-M\'editerran\'ee in France.  The current plan is to start constructing SPIP in 2018 for an 
implementation at TBL in late 2020.  Thanks to the 156\degr\ longitude shift of TBL with respect to CFHT, SPIP will be able 
to play a key role for monitoring stars with a higher time cadence and a denser coverage of the rotation cycles of the 
observed stars and of the orbital cycles of the detected planets as demonstrated by our latest coordinated observations of 
star/planet formation with ESPaDOnS and Narval \citep{Donati16, Donati17}.  

Last but not least, the feasibility of a SPIRou CubeSat working in parallel with SPIRou is also being studied.  The goal 
of this CubeSat would be to achieve continuous photometric monitoring in the JH bands at a precision of better than 1~mmag, 
over periods of up to 3~months and simultaneously with our main SPIRou observations.  Such complementary data would be quite 
useful to further characterize the activity of the host stars and to identify the transiting planets among all those 
that SPIRou will detect.

\begin{acknowledgement}
This paper is dedicated to the memory of Leslie Saddlemyer from NRC-H who passed away on 2017 Jan~09, and to that of Pierre Soler, 
director of OMP, who sadly left us shortly afterwards, on May~15.  Both played major roles in helping SPIRou come to life.  
Pierre was convinced that SPIRou was to be a key instrument for astronomy, and constantly supported the IRAP / OMP team managing 
the project in their quest for funding the instrument.  Without Leslie's strong dedication and his invaluable contribution to 
the AIT phases at IRAP / OMP since the spectrograph cryomechanics was delivered, reassembled and validated in Toulouse, SPIRou 
would not have reached the current stage of integration.  Both Leslie and Pierre will be remembered by the SPIRou team for their 
outstanding professional experience and their exceptional human qualities.  

The SPIRou team thanks all funding agencies in France (the IDEX initiatives in Toulouse and Marseille, DIM-ACAV in Paris, 
Labex OSUG@2020 in Grenoble, CNRS / INSU, Universit\'e de Toulouse Paul Sabatier and Universit\'e Grenoble-Alpes, R\'egion 
Occitanie / Pyr\'en\'ees-M\'editerran\'ee in Toulouse), Canada (CFI, NRC), Brazil (LNA), Switzerland (Geneva Observatory), 
Portugal (FCT), Taiwan (ASIAA) for their financial and / or manpower contribution to SPIRou.  We also thank the Board of 
CFHT for covering a significant fraction of SPIRou's construction costs and allocating human resources to the project.  

JFD further thanks Universit\'e Paul Sabatier, IRAP / OMP and CNRS / INSU for their strong and unfailing support for SPIRou,  
without which the project would not have been built.  
PF acknowledges support by Funda\c{c}\~ao para a Ci\^encia e a Tecnologia (FCT) through Investigador FCT contract of reference 
IF/01037/2013, and POPH/FSE (EC) by FEDER funding through the program ``Programa Operacional de Factores de Competitividade - COMPETE''. 
PF further acknowledges support from FCT in the form of an exploratory project of reference IF/01037/2013CP1191/CT0001.

\end{acknowledgement}

\bibliographystyle{spbasicHBexo}  
\bibliography{donati} 

\begin{thebibliography}{77}
\providecommand{\natexlab}[1]{#1}
\providecommand{\url}[1]{{#1}}
\providecommand{\urlprefix}{URL }
\expandafter\ifx\csname urlstyle\endcsname\relax
  \providecommand{\doi}[1]{DOI~\discretionary{}{}{}#1}\else
  \providecommand{\doi}{DOI~\discretionary{}{}{}\begingroup
  \urlstyle{rm}\Url}\fi
\providecommand{\eprint}[2][]{\url{#2}}

\bibitem[{{Allard} et~al.(2013){Allard}, {Homeier}, {Freytag},
  {Schaffenberger}, {}, and {Rajpurohit}}]{Allard13}
{Allard} F, {Homeier} D, {Freytag} B et~al. (2013) {Progress in modeling very
  low mass stars, brown dwarfs, and planetary mass objects.} Memorie della
  Societa Astronomica Italiana Supplementi 24:128

\bibitem[{{Andr{\'e}} et~al.(2009){Andr{\'e}}, {Basu}, and
  {Inutsuka}}]{Andre09}
{Andr{\'e}} P, {Basu} S {Inutsuka} S (2009) {The formation and evolution of
  prestellar cores}, Cambridge University Press, p 254

\bibitem[{{Anglada-Escud{\'e}} et~al.(2016){Anglada-Escud{\'e}}, {Amado},
  {Barnes}, {Berdi{\~n}as}, {Butler}, {Coleman}, {de La Cueva}, {Dreizler},
  {Endl}, {Giesers}, {Jeffers}, {Jenkins}, {Jones}, {Kiraga}, {K{\"u}rster},
  {L{\'o}pez-Gonz{\'a}lez}, {Marvin}, {Morales}, {Morin}, {Nelson}, {Ortiz},
  {Ofir}, {Paardekooper}, {Reiners}, {Rodr{\'{\i}}guez},
  {Rodr{\'{\i}}guez-L{\'o}pez}, {Sarmiento}, {Strachan}, {Tsapras}, {Tuomi},
  and {Zechmeister}}]{Anglada16}
{Anglada-Escud{\'e}} G, {Amado} PJ, {Barnes} J et~al. (2016) {A terrestrial
  planet candidate in a temperate orbit around Proxima Centauri}. \nat
  536:437--440

\bibitem[{{Artigau} et~al.(2009){Artigau}, {Bouchard}, {Doyon}, and
  {Lafreni{\`e}re}}]{Artigau09}
{Artigau} {\'E}, {Bouchard} S, {Doyon} R {Lafreni{\`e}re} D (2009) {Photometric
  Variability of the T2.5 Brown Dwarf SIMP J013656.5+093347: Evidence for
  Evolving Weather Patterns}. \apj 701:1534--1539

\bibitem[{{Artigau} et~al.(2014){Artigau}, {Kouach}, {Donati}, {Doyon},
  {Delfosse}, {Baratchart}, {Lacombe}, {Moutou}, {Rabou}, {Par{\`e}s},
  {Micheau}, {Thibault}, {Reshetov}, {Dubois}, {Hernandez}, {Vall{\'e}e},
  {Wang}, {Dolon}, {Pepe}, {Bouchy}, {Striebig}, {H{\'e}nault}, {Loop},
  {Saddlemyer}, {Barrick}, {Vermeulen}, {Dupieux}, {H{\'e}brard}, {Boisse},
  {Martioli}, {Alencar}, {do Nascimento}, and {Figueira}}]{Artigau14}
{Artigau} {\'E}, {Kouach} D, {Donati} JF et~al. (2014) {SPIRou: the
  near-infrared spectropolarimeter/high-precision velocimeter for the
  Canada-France-Hawaii telescope}. In: Ground-based and Airborne
  Instrumentation for Astronomy V, \procspie, vol 9147, p 914715,
  \doi{10.1117/12.2055663}

\bibitem[{{Baraffe} and {Chabrier}(2010)}]{Baraffe10}
{Baraffe} I {Chabrier} G (2010) {Effect of episodic accretion on the structure
  and the lithium depletion of low-mass stars and planet-hosting stars}. \aap
  521:A44

\bibitem[{{Barrick} et~al.(2012){Barrick}, {Vermeulen}, {Baratchart},
  {Reshetov}, {Wang}, {Dolon}, {Hernandez}, {Pepe}, {Bouchy}, {Dunn},
  {Dupieux}, {Gallou}, {Larrieu}, {Fonteneau}, {Moreau}, {Wildi}, {Par{\`e}s},
  {Thomas}, {Yan}, {Doyon}, {Donati}, {Vall{\'e}e}, {Artigau}, {Delfosse},
  {Rabou}, {Thibault}, {Kouach}, and {Loop}}]{Barrick12}
{Barrick} GA, {Vermeulen} T, {Baratchart} S et~al. (2012) {SPIRou @ CFHT:
  design of the instrument control system}. In: Software and
  Cyberinfrastructure for Astronomy II, \procspie, vol 8451, p 84513J,
  \doi{10.1117/12.926392}

\bibitem[{{Baruteau} et~al.(2014){Baruteau}, {Crida}, {Paardekooper}, {Masset},
  {Guilet}, {Bitsch}, {Nelson}, {Kley}, and {Papaloizou}}]{Baruteau14}
{Baruteau} C, {Crida} A, {Paardekooper} SJ et~al. (2014) {Planet-Disk
  Interactions and Early Evolution of Planetary Systems}. Protostars and
  Planets VI pp 667--689

\bibitem[{{Berta-Thompson} et~al.(2015){Berta-Thompson}, {Irwin},
  {Charbonneau}, {Newton}, {Dittmann}, {Astudillo-Defru}, {Bonfils}, {Gillon},
  {Jehin}, {Stark}, {Stalder}, {Bouchy}, {Delfosse}, {Forveille}, {Lovis},
  {Mayor}, {Neves}, {Pepe}, {Santos}, {Udry}, and {W{\"u}nsche}}]{Berta15}
{Berta-Thompson} ZK, {Irwin} J, {Charbonneau} D et~al. (2015) {A rocky planet
  transiting a nearby low-mass star}. \nat 527:204--207

\bibitem[{{Blinova} et~al.(2016){Blinova}, {Romanova}, and
  {Lovelace}}]{Blinova16}
{Blinova} AA, {Romanova} MM {Lovelace} RVE (2016) {Boundary between stable and
  unstable regimes of accretion. Ordered and chaotic unstable regimes}. \mnras
  459:2354--2369

\bibitem[{{Boisse} et~al.(2016){Boisse}, {Perruchot}, {Bouchy}, {Dolon},
  {Moreau}, {Sottile}, and {Wildi}}]{Boisse16}
{Boisse} I, {Perruchot} S, {Bouchy} F et~al. (2016) {A calibration unit for the
  near-infrared spectropolarimeter SPIRou}. In: Society of Photo-Optical
  Instrumentation Engineers (SPIE) Conference Series, \procspie, vol 9908, p
  990868, \doi{10.1117/12.2231678}

\bibitem[{{Bonfils} et~al.(2013){Bonfils}, {Delfosse}, {Udry}, {Forveille},
  {Mayor}, {Perrier}, {Bouchy}, {Gillon}, {Lovis}, {Pepe}, {Queloz}, {Santos},
  {S{\'e}gransan}, and {Bertaux}}]{Bonfils13}
{Bonfils} X, {Delfosse} X, {Udry} S et~al. (2013) {The HARPS search for
  southern extra-solar planets. XXXI. The M-dwarf sample}. \aap 549:A109

\bibitem[{{Bouvier} et~al.(2007){Bouvier}, {Alencar}, {Harries}, {Johns-Krull},
  and {Romanova}}]{Bouvier07}
{Bouvier} J, {Alencar} SHP, {Harries} TJ, {Johns-Krull} CM {Romanova} MM (2007)
  {Magnetospheric Accretion in Classical T Tauri Stars}. In: {Reipurth} B,
  {Jewitt} D {Keil} K (eds) Protostars and Planets V, pp 479--494

\bibitem[{{Bower} et~al.(2016){Bower}, {Loinard}, {Dzib}, {Galli},
  {Ortiz-Le{\'o}n}, {Moutou}, and {Donati}}]{Bower16}
{Bower} GC, {Loinard} L, {Dzib} S et~al. (2016) {Variable Radio Emission from
  the Young Stellar Host of a Hot Jupiter}. \apj 830:107

\bibitem[{{Brogi} et~al.(2012){Brogi}, {Snellen}, {de Kok}, {Albrecht},
  {Birkby}, and {de Mooij}}]{Brogi12}
{Brogi} M, {Snellen} IAG, {de Kok} RJ et~al. (2012) {The signature of orbital
  motion from the dayside of the planet {$\tau$} Bo{\"o}tis b}. \nat
  486:502--504

\bibitem[{{Carmona} et~al.(2013){Carmona}, {Bouvier}, and
  {Delfosse}}]{Carmona13}
{Carmona} A, {Bouvier} J {Delfosse} X (2013) {Perspectives for the study of gas
  in protoplanetary disks and accretion/ejection phenomena in young stars with
  the near-IR spectrograph SPIROU at the CFHT}. In: {Cambresy} L, {Martins} F,
  {Nuss} E {Palacios} A (eds) SF2A-2013: Proceedings of the Annual meeting of
  the French Society of Astronomy and Astrophysics, pp 493--495

\bibitem[{{Cody} et~al.(2014){Cody}, {Stauffer}, {Baglin}, {Micela}, {Rebull},
  {Flaccomio}, {Morales-Calder{\'o}n}, {Aigrain}, {Bouvier}, {Hillenbrand},
  {Gutermuth}, {Song}, {Turner}, {Alencar}, {Zwintz}, {Plavchan}, {Carpenter},
  {Findeisen}, {Carey}, {Terebey}, {Hartmann}, {Calvet}, {Teixeira}, {Vrba},
  {Wolk}, {Covey}, {Poppenhaeger}, {G{\"u}nther}, {Forbrich}, {Whitney},
  {Affer}, {Herbst}, {Hora}, {Barrado}, {Holtzman}, {Marchis}, {Wood},
  {Medeiros Guimar{\~a}es}, {Lillo Box}, {Gillen}, {McQuillan}, {Espaillat},
  {Allen}, {D'Alessio}, and {Favata}}]{Cody14}
{Cody} AM, {Stauffer} J, {Baglin} A et~al. (2014) {CSI 2264: Simultaneous
  Optical and Infrared Light Curves of Young Disk-bearing Stars in NGC 2264
  with CoRoT and Spitzer. Evidence for Multiple Origins of Variability}. \aj
  147:82

\bibitem[{{Crossfield} et~al.(2014){Crossfield}, {Biller}, {Schlieder},
  {Deacon}, {Bonnefoy}, {Homeier}, {Allard}, {Buenzli}, {Henning}, {Brandner},
  {Goldman}, and {Kopytova}}]{Crossfield14}
{Crossfield} IJM, {Biller} B, {Schlieder} JE et~al. (2014) {A global cloud map
  of the nearest known brown dwarf}. \nat 505:654--656

\bibitem[{{David} et~al.(2016){David}, {Hillenbrand}, {Petigura}, {Carpenter},
  {Crossfield}, {Hinkley}, {Ciardi}, {Howard}, {Isaacson}, {Cody}, {Schlieder},
  {Beichman}, and {Barenfeld}}]{David16}
{David} TJ, {Hillenbrand} LA, {Petigura} EA et~al. (2016) {A Neptune-sized
  transiting planet closely orbiting a 5-10-million-year-old star}. \nat
  534:658--661

\bibitem[{{Davies} et~al.(2014){Davies}, {Gregory}, and {Greaves}}]{Davies14}
{Davies} CL, {Gregory} SG {Greaves} JS (2014) {Accretion discs as regulators of
  stellar angular momentum evolution in the ONC and Taurus-Auriga}. \mnras
  444:1157--1176

\bibitem[{{Delfosse} et~al.(2013){Delfosse}, {Donati}, {Kouach}, {H{\'e}brard},
  {Doyon}, {Artigau}, {Bouchy}, {Boisse}, {Brun}, {Hennebelle}, {Widemann},
  {Bouvier}, {Bonfils}, {Morin}, {Moutou}, {Pepe}, {Udry}, {do Nascimento},
  {Alencar}, {Castilho}, {Martioli}, {Wang}, {Figueira}, and
  {Santos}}]{Delfosse13}
{Delfosse} X, {Donati} JF, {Kouach} D et~al. (2013) {World-leading science with
  SPIRou - The nIR spectropolarimeter / high-precision velocimeter for CFHT}.
  In: {Cambresy} L, {Martins} F, {Nuss} E {Palacios} A (eds) SF2A-2013:
  Proceedings of the Annual meeting of the French Society of Astronomy and
  Astrophysics, pp 497--508

\bibitem[{{Donati} et~al.(2010){Donati}, {Skelly}, {Bouvier}, {Jardine},
  {Gregory}, {Morin}, {Hussain}, {Dougados}, {M{\'e}nard}, and
  {Unruh}}]{Donati10}
{Donati} J, {Skelly} MB, {Bouvier} J et~al. (2010) {Complex magnetic topology
  and strong differential rotation on the low-mass T Tauri star V2247 Oph}.
  \mnras 402:1426--1436

\bibitem[{{Donati}(2003)}]{Donati03}
{Donati} JF (2003) {ESPaDOnS: An Echelle SpectroPolarimetric Device for the
  Observation of Stars at CFHT}. In: {Trujillo-Bueno} J {Sanchez Almeida} J
  (eds) Astronomical Society of the Pacific Conference Series, Astronomical
  Society of the Pacific Conference Series, vol 307, p~41

\bibitem[{{Donati} et~al.(1997){Donati}, {Semel}, {Carter}, {Rees}, and
  {Collier Cameron}}]{Donati97b}
{Donati} JF, {Semel} M, {Carter} BD, {Rees} DE {Collier Cameron} A (1997)
  {Spectropolarimetric observations of active stars}. \mnras 291:658

\bibitem[{{Donati} et~al.(2005){Donati}, {Paletou}, {Bouvier}, and
  {Ferreira}}]{Donati05}
{Donati} JF, {Paletou} F, {Bouvier} J {Ferreira} J (2005) {Direct detection of
  a magnetic field in the innermost regions of an accretion disk}. \nat
  438:466--469

\bibitem[{{Donati} et~al.(2006){Donati}, {Catala}, {Landstreet}, and
  {Petit}}]{Donati06d}
{Donati} JF, {Catala} C, {Landstreet} JD {Petit} P (2006) {ESPaDOnS: The New
  Generation Stellar Spectro-Polarimeter. Performances and First Results}. In:
  {Casini} R {Lites} BW (eds) Astronomical Society of the Pacific Conference
  Series, Astronomical Society of the Pacific Conference Series, vol 358, p 362

\bibitem[{{Donati} et~al.(2012){Donati}, {Gregory}, {Alencar}, {Hussain},
  {Bouvier}, {Dougados}, {Jardine}, {M{\'e}nard}, and {Romanova}}]{Donati12}
{Donati} JF, {Gregory} SG, {Alencar} SHP et~al. (2012) {Magnetometry of the
  classical T Tauri star GQ Lup: non-stationary dynamos and spin evolution of
  young Suns}. \mnras 425:2948--2963

\bibitem[{{Donati} et~al.(2013){Donati}, {Gregory}, {Alencar}, {Hussain},
  {Bouvier}, {Jardine}, {M{\'e}nard}, {Dougados}, {Romanova}, and {MaPP
  Collaboration}}]{Donati13}
{Donati} JF, {Gregory} SG, {Alencar} SHP et~al. (2013) {Magnetospheric
  accretion on the fully convective classical T Tauri star DN Tau}. \mnras
  436:881--897

\bibitem[{{Donati} et~al.(2014){Donati}, {H{\'e}brard}, {Hussain}, {Moutou},
  {Grankin}, {Boisse}, {Morin}, {Gregory}, {Vidotto}, {Bouvier}, {Alencar}, and
  {et al.,}}]{Donati14}
{Donati} JF, {H{\'e}brard} E, {Hussain} G et~al. (2014) {Modelling the magnetic
  activity and filtering radial velocity curves of young Suns : the weak-line T
  Tauri star LkCa 4}. \mnras 444:3220--3229

\bibitem[{{Donati} et~al.(2016){Donati}, {Moutou}, {Malo}, {Baruteau}, {Yu},
  {H{\'e}brard}, {Hussain}, {Alencar}, {M{\'e}nard}, {Bouvier}, {Petit},
  {Takami}, {Doyon}, and {Cameron}}]{Donati16}
{Donati} JF, {Moutou} C, {Malo} L et~al. (2016) {A hot Jupiter orbiting a
  2-million-year-old solar-mass T Tauri star}. \nat 534:662--666

\bibitem[{{Donati} et~al.(2017){Donati}, {Yu}, {Moutou}, {Cameron}, {Malo},
  {Grankin}, {H{\'e}brard}, {Hussain}, {Vidotto}, {Alencar}, {Haywood},
  {Bouvier}, {Petit}, {Takami}, {Herczeg}, {Gregory}, {Jardine}, and
  {Morin}}]{Donati17}
{Donati} JF, {Yu} L, {Moutou} C et~al. (2017) {The hot Jupiter of the
  magnetically active weak-line T Tauri star V830 Tau}. \mnras 465:3343--3360

\bibitem[{{Dressing} and {Charbonneau}(2015)}]{Dressing15}
{Dressing} CD {Charbonneau} D (2015) {The Occurrence of Potentially Habitable
  Planets Orbiting M Dwarfs Estimated from the Full Kepler Dataset and an
  Empirical Measurement of the Detection Sensitivity}. \apj 807:45

\bibitem[{{Feiden}(2016)}]{Feiden16}
{Feiden} GA (2016) {Magnetic inhibition of convection and the fundamental
  properties of low-mass stars. III. A consistent 10 Myr age for the Upper
  Scorpius OB association}. \aap 593:A99

\bibitem[{{Gaidos} and {Selsis}(2007)}]{Gaidos07}
{Gaidos} E {Selsis} F (2007) {From Protoplanets to Protolife: The Emergence and
  Maintenance of Life}. Protostars and Planets V pp 929--944

\bibitem[{{Gaidos} et~al.(2016){Gaidos}, {Mann}, {Kraus}, and
  {Ireland}}]{Gaidos16}
{Gaidos} E, {Mann} AW, {Kraus} AL {Ireland} M (2016) {They are small worlds
  after all: revised properties of Kepler M dwarf stars and their planets}.
  \mnras 457:2877--2899

\bibitem[{{Gomes da Silva} et~al.(2012){Gomes da Silva}, {Santos}, {Bonfils},
  {Delfosse}, {Forveille}, {Udry}, {Dumusque}, and {Lovis}}]{Gomes12}
{Gomes da Silva} J, {Santos} NC, {Bonfils} X et~al. (2012) {Long-term magnetic
  activity of a sample of M-dwarf stars from the HARPS program . II. Activity
  and radial velocity}. \aap 541:A9

\bibitem[{{Gregory} et~al.(2012){Gregory}, {Donati}, {Morin}, {Hussain},
  {Mayne}, {Hillenbrand}, and {Jardine}}]{Gregory12}
{Gregory} SG, {Donati} JF, {Morin} J et~al. (2012) {Can We Predict the Global
  Magnetic Topology of a Pre-main-sequence Star from Its Position in the
  Hertzsprung-Russell Diagram?} \apj 755:97

\bibitem[{{G{\"u}del} et~al.(2014){G{\"u}del}, {Dvorak}, {Erkaev}, {Kasting},
  {Khodachenko}, {Lammer}, {Pilat-Lohinger}, {Rauer}, {Ribas}, and
  {Wood}}]{Gudel14}
{G{\"u}del} M, {Dvorak} R, {Erkaev} N et~al. (2014) {Astrophysical Conditions
  for Planetary Habitability}. Protostars and Planets VI pp 883--906

\bibitem[{{H{\'e}brard} et~al.(2016){H{\'e}brard}, {Donati}, {Delfosse},
  {Morin}, {Moutou}, and {Boisse}}]{Hebrard16}
{H{\'e}brard} {\'E}M, {Donati} JF, {Delfosse} X et~al. (2016) {Modelling the RV
  jitter of early-M dwarfs using tomographic imaging}. \mnras 461:1465--1497

\bibitem[{{Johansen}(2009)}]{Johansen09}
{Johansen} A (2009) {The role of magnetic fields for planetary formation}. In:
  {Strassmeier} KG, {Kosovichev} AG {Beckman} JE (eds) Cosmic Magnetic Fields:
  From Planets, to Stars and Galaxies, IAU Symposium, vol 259, pp 249--258,
  \doi{10.1017/S1743921309030592}

\bibitem[{{Johns-Krull} et~al.(2009){Johns-Krull}, {Greene}, {Doppmann}, and
  {Covey}}]{Johns09}
{Johns-Krull} CM, {Greene} TP, {Doppmann} GW {Covey} KR (2009) {First Magnetic
  Field Detection on a Class I Protostar}. \apj 700:1440--1448

\bibitem[{{Lin} et~al.(1996){Lin}, {Bodenheimer}, and {Richardson}}]{Lin96}
{Lin} DNC, {Bodenheimer} P {Richardson} DC (1996) {Orbital migration of the
  planetary companion of 51 Pegasi to its present location}. \nat 380:606--607

\bibitem[{{Lissauer} et~al.(2014){Lissauer}, {Dawson}, and
  {Tremaine}}]{Lissauer14}
{Lissauer} JJ, {Dawson} RI {Tremaine} S (2014) {Advances in exoplanet science
  from Kepler}. \nat 513:336--344

\bibitem[{{Machado} et~al.(2014){Machado}, {Widemann}, {Luz}, and
  {Peralta}}]{Machado14}
{Machado} P, {Widemann} T, {Luz} D {Peralta} J (2014) {Wind circulation regimes
  at Venus' cloud tops: Ground-based Doppler velocimetry using CFHT/ESPaDOnS
  and comparison with simultaneous cloud tracking measurements using VEx/VIRTIS
  in February 2011}. \icarus 243:249--263

\bibitem[{{Machado} et~al.(2017){Machado}, {Widemann}, {Peralta}, {Gon{\c
  c}alves}, {Donati}, and {Luz}}]{Machado17}
{Machado} P, {Widemann} T, {Peralta} J et~al. (2017) {Venus cloud-tracked and
  doppler velocimetry winds from CFHT/ESPaDOnS and Venus Express/VIRTIS in
  April 2014}. \icarus 285:8--26

\bibitem[{{Maury} et~al.(2010){Maury}, {Andr{\'e}}, {Hennebelle}, {Motte},
  {Stamatellos}, {Bate}, {Belloche}, {Duch{\^e}ne}, and {Whitworth}}]{Maury10}
{Maury} AJ, {Andr{\'e}} P, {Hennebelle} P et~al. (2010) {Toward understanding
  the formation of multiple systems. A pilot IRAM-PdBI survey of Class 0
  objects}. \aap 512:A40

\bibitem[{{Micheau} et~al.(2012){Micheau}, {Bouchy}, {Pepe}, {Chazelas},
  {Kouach}, {Par{\`e}s}, {Donati}, {Barrick}, {Rabou}, {Thibault},
  {Saddlemyer}, {Perruchot}, {Delfosse}, {Striebig}, {Gallou}, {Loop}, and
  {Pazder}}]{Micheau12}
{Micheau} Y, {Bouchy} F, {Pepe} F et~al. (2012) {SPIRou @ CFHT: fiber links and
  pupil slicer}. In: Ground-based and Airborne Instrumentation for Astronomy
  IV, \procspie, vol 8446, p 84462R, \doi{10.1117/12.926084}

\bibitem[{{Micheau} et~al.(2015){Micheau}, {Bouy{\'e}}, {Parisot}, and
  {Kouach}}]{Micheau15}
{Micheau} Y, {Bouy{\'e}} M, {Parisot} J {Kouach} D (2015) {Fluoride fiber
  thermal emission study for SPIRou @ CFHT}. In: Techniques and Instrumentation
  for Detection of Exoplanets VII, \procspie, vol 9605, p 96051Q,
  \doi{10.1117/12.2185188}

\bibitem[{{Morin} et~al.(2008){Morin}, {Donati}, {Petit}, {Delfosse},
  {Forveille}, {Albert}, {Auri{\`e}re}, {Cabanac}, {Dintrans}, {Fares},
  {Gastine}, {Jardine}, {Ligni{\`e}res}, {Paletou}, {Ramirez Velez}, and
  {Th{\'e}ado}}]{Morin08b}
{Morin} J, {Donati} JF, {Petit} P et~al. (2008) {Large-scale magnetic
  topologies of mid M dwarfs}. \mnras 390:567--581

\bibitem[{{Morin} et~al.(2010){Morin}, {Donati}, {Petit}, {Delfosse},
  {Forveille}, and {Jardine}}]{Morin10}
{Morin} J, {Donati} J, {Petit} P et~al. (2010) {Large-scale magnetic topologies
  of late M dwarfs}. \mnras 407:2269--2286

\bibitem[{{Moutou} et~al.(2015){Moutou}, {Boisse}, {H{\'e}brard},
  {H{\'e}brard}, {Donati}, {Delfosse}, and {Kouach}}]{Moutou15}
{Moutou} C, {Boisse} I, {H{\'e}brard} G et~al. (2015) {SPIRou: a
  spectropolarimeter for the CFHT}. In: {Martins} F, {Boissier} S, {Buat} V,
  {Cambr{\'e}sy} L {Petit} P (eds) SF2A-2015: Proceedings of the Annual meeting
  of the French Society of Astronomy and Astrophysics, pp 205--212

\bibitem[{{Muirhead} et~al.(2012){Muirhead}, {Johnson}, {Apps}, {Carter},
  {Morton}, {Fabrycky}, {Pineda}, {Bottom}, {Rojas-Ayala}, {Schlawin},
  {Hamren}, {Covey}, {Crepp}, {Stassun}, {Pepper}, {Hebb}, {Kirby}, {Howard},
  {Isaacson}, {Marcy}, {Levitan}, {Diaz-Santos}, {Armus}, and
  {Lloyd}}]{Muirhead12}
{Muirhead} PS, {Johnson} JA, {Apps} K et~al. (2012) {Characterizing the Cool
  KOIs. III. KOI 961: A Small Star with Large Proper Motion and Three Small
  Planets}. \apj 747:144

\bibitem[{{Muirhead} et~al.(2015){Muirhead}, {Mann}, {Vanderburg}, {Morton},
  {Kraus}, {Ireland}, {Swift}, {Feiden}, {Gaidos}, and {Gazak}}]{Muirhead15}
{Muirhead} PS, {Mann} AW, {Vanderburg} A et~al. (2015) {Kepler-445, Kepler-446
  and the Occurrence of Compact Multiples Orbiting Mid-M Dwarf Stars}. \apj
  801:18

\bibitem[{{Newton} et~al.(2016){Newton}, {Irwin}, {Charbonneau},
  {Berta-Thompson}, and {Dittmann}}]{Newton16}
{Newton} ER, {Irwin} J, {Charbonneau} D, {Berta-Thompson} ZK {Dittmann} JA
  (2016) {The Impact of Stellar Rotation on the Detectability of Habitable
  Planets around M Dwarfs}. \apjl 821:L19

\bibitem[{{Par{\`e}s} et~al.(2012){Par{\`e}s}, {Donati}, {Dupieux}, {Gharsa},
  {Micheau}, {Bouye}, {Dubois}, {Gallou}, {Kouach}, {Barrick}, and
  {Wang}}]{Pares12}
{Par{\`e}s} L, {Donati} JF, {Dupieux} M et~al. (2012) {Front end of the SPIRou
  spectropolarimeter for Canada-France Hawaii Telescope}. In: Ground-based and
  Airborne Instrumentation for Astronomy IV, \procspie, vol 8446, p 84462E,
  \doi{10.1117/12.925410}

\bibitem[{{Passegger} et~al.(2016){Passegger}, {Wende-von Berg}, and
  {Reiners}}]{Passegger16}
{Passegger} VM, {Wende-von Berg} S {Reiners} A (2016) {Fundamental M-dwarf
  parameters from high-resolution spectra using PHOENIX ACES models. I.
  Parameter accuracy and benchmark stars}. \aap 587:A19

\bibitem[{{Pepe} et~al.(2003){Pepe}, {Rupprecht}, {Avila}, {Balestra},
  {Bouchy}, {Cavadore}, {Eckert}, {Fleury}, {Gillotte}, {Gojak}, {Guzman},
  {Kohler}, {Lizon}, {Mayor}, {Megevand}, {Queloz}, {Sosnowska}, {Udry}, and
  {Weilenmann}}]{Pepe03}
{Pepe} F, {Rupprecht} G, {Avila} G et~al. (2003) {Performance verification of
  HARPS: first laboratory results}. In: {Iye} M {Moorwood} AFM (eds) Instrument
  Design and Performance for Optical/Infrared Ground-based Telescopes,
  \procspie, vol 4841, pp 1045--1056, \doi{10.1117/12.460777}

\bibitem[{{Rajpurohit} et~al.(2013){Rajpurohit}, {Reyl{\'e}}, {Allard},
  {Homeier}, {Schultheis}, {Bessell}, and {Robin}}]{Rajpurohit13}
{Rajpurohit} AS, {Reyl{\'e}} C, {Allard} F et~al. (2013) {The effective
  temperature scale of M dwarfs}. \aap 556:A15

\bibitem[{{Reggiani} et~al.(2016){Reggiani}, {Mel{\'e}ndez}, {Yong},
  {Ram{\'{\i}}rez}, and {Asplund}}]{Reggiani16}
{Reggiani} H, {Mel{\'e}ndez} J, {Yong} D, {Ram{\'{\i}}rez} I {Asplund} M (2016)
  {First high-precision differential abundance analysis of extremely metal-poor
  stars}. \aap 586:A67

\bibitem[{{Reshetov} et~al.(2012){Reshetov}, {Herriot}, {Thibault},
  {D{\'e}saulniers}, {Saddlemyer}, and {Loop}}]{Reshetov12}
{Reshetov} V, {Herriot} G, {Thibault} S et~al. (2012) {Cryogenic mechanical
  design: SPIROU spectrograph}. In: Ground-based and Airborne Instrumentation
  for Astronomy IV, \procspie, vol 8446, p 84464E, \doi{10.1117/12.927442}

\bibitem[{{Romanova} and {Lovelace}(2006)}]{Romanova06}
{Romanova} MM {Lovelace} RVE (2006) {The Magnetospheric Gap and the
  Accumulation of Giant Planets Close to a Star}. \apjl 645:L73--L76

\bibitem[{{Romanova} et~al.(2004){Romanova}, {Ustyugova}, {Koldoba}, and
  {Lovelace}}]{Romanova04}
{Romanova} MM, {Ustyugova} GV, {Koldoba} AV {Lovelace} RVE (2004) {The
  Propeller Regime of Disk Accretion to a Rapidly Rotating Magnetized Star}.
  \apjl 616:L151--L154

\bibitem[{{Romanova} et~al.(2008){Romanova}, {Kulkarni}, and
  {Lovelace}}]{Romanova08}
{Romanova} MM, {Kulkarni} AK {Lovelace} RVE (2008) {Unstable Disk Accretion
  onto Magnetized Stars: First Global Three-dimensional Magnetohydrodynamic
  Simulations}. \apjl 673:L171

\bibitem[{{Romanova} et~al.(2011){Romanova}, {Long}, {Lamb}, {Kulkarni}, and
  {Donati}}]{Romanova11}
{Romanova} MM, {Long} M, {Lamb} FK, {Kulkarni} AK {Donati} J (2011) {Global 3D
  simulations of disc accretion on to the classical T Tauri star V2129 Oph}.
  \mnras 411:915--928

\bibitem[{{Rupprecht} et~al.(2004){Rupprecht}, {Pepe}, {Mayor}, {Queloz},
  {Bouchy}, {Avila}, {Benz}, {Bertaux}, {Bonfils}, {Dall}, {Delabre}, {Dekker},
  {Eckert}, {Fleury}, {Gilliotte}, {Gojak}, {Guzman}, {Kohler}, {Lizon}, {Lo
  Curto}, {Longinotti}, {Lovis}, {Megevand}, {Pasquini}, {Reyes}, {Sivan},
  {Sosnowska}, {Soto}, {Udry}, {Van Kesteren}, {Weber}, and
  {Weilenmann}}]{Rupprecht04}
{Rupprecht} G, {Pepe} F, {Mayor} M et~al. (2004) {The exoplanet hunter HARPS:
  performance and first results}. In: {Moorwood} AFM {Iye} M (eds) Ground-based
  Instrumentation for Astronomy, \procspie, vol 5492, pp 148--159,
  \doi{10.1117/12.551267}

\bibitem[{{Santerne} et~al.(2013){Santerne}, {Donati}, {Doyon}, {Delfosse},
  {Artigau}, {Boisse}, {Bonfils}, {Bouchy}, {H{\'e}brard}, {Moutou}, and
  {Udry}}]{Santerne13}
{Santerne} A, {Donati} JF, {Doyon} R et~al. (2013) {Characterizing small
  planets transiting small stars with SPIRou}. In: {Cambresy} L, {Martins} F,
  {Nuss} E {Palacios} A (eds) SF2A-2013: Proceedings of the Annual meeting of
  the French Society of Astronomy and Astrophysics, pp 509--514

\bibitem[{{Shu} et~al.(2007){Shu}, {Galli}, {Lizano}, {Glassgold}, and
  {Diamond}}]{Shu07}
{Shu} FH, {Galli} D, {Lizano} S, {Glassgold} AE {Diamond} PH (2007) {Mean Field
  Magnetohydrodynamics of Accretion Disks}. \apj 665:535--553

\bibitem[{{Skelly} et~al.(2010){Skelly}, {Donati}, {Bouvier}, {Grankin},
  {Unruh}, {Artemenko}, and {Petrov}}]{Skelly10}
{Skelly} MB, {Donati} JF, {Bouvier} J et~al. (2010) {Dynamo processes in the T
  Tauri star V410 Tau}. \mnras 403:159--169

\bibitem[{{Snellen} et~al.(2010){Snellen}, {de Kok}, {de Mooij}, and
  {Albrecht}}]{Snellen10}
{Snellen} IAG, {de Kok} RJ, {de Mooij} EJW {Albrecht} S (2010) {The orbital
  motion, absolute mass and high-altitude winds of exoplanet HD209458b}. \nat
  465:1049--1051

\bibitem[{{Sousa} et~al.(2016){Sousa}, {Alencar}, {Bouvier}, {Stauffer},
  {Venuti}, {Hillenbrand}, {Cody}, {Teixeira}, {Guimar{\~a}es}, {McGinnis},
  {Rebull}, {Flaccomio}, {F{\"u}r{\'e}sz}, {Micela}, and {Gameiro}}]{Sousa16}
{Sousa} AP, {Alencar} SHP, {Bouvier} J et~al. (2016) {CSI 2264: Accretion
  process in classical T Tauri stars in the young cluster NGC 2264}. \aap
  586:A47

\bibitem[{{Strugarek} et~al.(2015){Strugarek}, {Brun}, {Matt}, and
  {R{\'e}ville}}]{Strugarek15}
{Strugarek} A, {Brun} AS, {Matt} SP {R{\'e}ville} V (2015) {Magnetic Games
  between a Planet and Its Host Star: The Key Role of Topology}. \apj 815:111

\bibitem[{{Sullivan} et~al.(2015){Sullivan}, {Winn}, {Berta-Thompson},
  {Charbonneau}, {Deming}, {Dressing}, {Latham}, {Levine}, {McCullough},
  {Morton}, {Ricker}, {Vanderspek}, and {Woods}}]{Sullivan15}
{Sullivan} PW, {Winn} JN, {Berta-Thompson} ZK et~al. (2015) {The Transiting
  Exoplanet Survey Satellite: Simulations of Planet Detections and
  Astrophysical False Positives}. \apj 809:77

\bibitem[{{Thibault} et~al.(2012){Thibault}, {Rabou}, {Donati}, {Desaulniers},
  {Dallaire}, {Artigau}, {Pepe}, {Micheau}, {Vall{\'e}e}, {Pepe}, {Barrick},
  {Reshetov}, {Hernandez}, {Saddlemyer}, {Pazder}, {Par{\`e}s}, {Doyon},
  {Delfosse}, {Kouach}, and {Loop}}]{Thibault12}
{Thibault} S, {Rabou} P, {Donati} JF et~al. (2012) {SPIRou @ CFHT: spectrograph
  optical design}. In: Ground-based and Airborne Instrumentation for Astronomy
  IV, \procspie, vol 8446, p 844630, \doi{10.1117/12.926697}

\bibitem[{{Vidotto} and {Donati}(2017)}]{Vidotto17}
{Vidotto} AA {Donati} JF (2017) {Predicting radio emission from the newborn hot
  Jupiter V830~Tau~b}. \aap submitted

\bibitem[{{Vidotto} et~al.(2013){Vidotto}, {Jardine}, {Morin}, {Donati},
  {Lang}, and {Russell}}]{Vidotto13}
{Vidotto} AA, {Jardine} M, {Morin} J et~al. (2013) {Effects of M dwarf magnetic
  fields on potentially habitable planets}. \aap 557:A67

\bibitem[{{Yu} et~al.(2017){Yu}, {Donati}, {H{\'e}brard}, {Moutou}, {Malo},
  {Grankin}, {Hussain}, {Collier Cameron}, {Vidotto}, {Baruteau}, {Alencar},
  {Bouvier}, {Petit}, {Takami}, {Herczeg}, {Gregory}, {Jardine}, {Morin},
  {M{\'e}nard}, and {Matysse Collaboration}}]{Yu17}
{Yu} L, {Donati} JF, {H{\'e}brard} EM et~al. (2017) {A hot Jupiter around the
  very active weak-line T Tauri star TAP 26}. \mnras

\bibitem[{{Zanni} and {Ferreira}(2013)}]{Zanni13}
{Zanni} C {Ferreira} J (2013) {MHD simulations of accretion onto a dipolar
  magnetosphere. II. Magnetospheric ejections and stellar spin-down}. \aap
  550:A99

\end{thebibliography}

\end{document}